\UseRawInputEncoding

\documentclass[preprint,12pt]{elsarticle}

\usepackage{amsmath}
\usepackage{amssymb}
\usepackage{amsfonts}
\usepackage[english]{babel}
\usepackage{csquotes}
\usepackage{microtype}
\usepackage{siunitx}
\usepackage{listings}
\usepackage{xcolor}
\usepackage{booktabs}
\usepackage{mathtools}
\usepackage{fancyvrb}
\usepackage{enumerate}
\usepackage{longtable}
\usepackage{multirow}
\usepackage{siunitx}
\usepackage{verbatim}
\usepackage{bbold}
\usepackage{caption}
\usepackage{hyperref}
\usepackage{cleveref}
\usepackage{dirtree}
\usepackage{array}
\usepackage{colortbl}
\usepackage{float}
\usepackage{tablefootnote}
\usepackage[T1]{fontenc}
\definecolor{mygray}{HTML}{e8e8e8}

\definecolor{codegreen}{rgb}{0,0.6,0}
\definecolor{codegray}{rgb}{0.5,0.5,0.5}
\definecolor{codepurple}{rgb}{0.58,0,0.82}
\definecolor{backcolour}{rgb}{0.95,0.95,0.92}
\definecolor{magenta}{rgb}{1.0, 0., 1.0}

\definecolor{codegreen}{rgb}{0,0.6,0}
\definecolor{codegray}{rgb}{0.5,0.5,0.5}
\definecolor{codepurple}{rgb}{0.58,0,0.82}
\definecolor{backcolour}{rgb}{0.95,0.95,0.92}

\lstdefinestyle{mystyle}{
    backgroundcolor=\color{backcolour},   
    commentstyle=\color{codegreen},
    keywordstyle=\color{magenta},
    numberstyle=\tiny\color{codegray},
    stringstyle=\color{codepurple},
    basicstyle=\ttfamily\footnotesize,
    breakatwhitespace=false,         
    breaklines=true,                 
    captionpos=b,                    
    keepspaces=true,                 
    numbers=left,                    
    numbersep=5pt,                  
    showspaces=false,                
    showstringspaces=false,
    showtabs=false,                  
    tabsize=2
}

\journal{Journal of Quantitative Spectroscopy and Radiative Transfer}

\begin{document}

\begin{frontmatter}



\title{T-matrix representation of optical scattering response: Suggestion for a data format}


\author[KIT]{Nigar Asadova}
\author[EPFL]{Karim Achouri}
\author[Aalto]{Kristian Arjas}
\author[Wellington]{Baptiste Augui\'e}
\author[TUHH]{Roland Aydin}
\author[PaulPascal]{Alexandre Baron}
\author[KIT]{Dominik Beutel}
\author[PTB]{Bernd Bodermann}
\author[KIT]{Kaoutar Boussaoud}
\author[ZIB]{Sven Burger}
\author[POSTECH]{Minseok Choi}
\author[UniWarsaw]{Krzysztof M. Czajkowski}
\author[Hannover]{Andrey B. Evlyukhin}
\author[Wellington]{Atefeh Fazel-Najafabadi}
\author[KIT]{Ivan Fernandez-Corbaton}
\author[KIT]{Puneet Garg}
\author[Wien]{David Globosits}
\author[Graz]{Ulrich Hohenester}
\author[POSTECH]{Hongyoon Kim}
\author[POSTECH]{Seokwoo Kim}
\author[LP2N]{Philippe Lalanne}
\author[Wellington]{Eric C. Le Ru}
\author[KIT]{J\"org Meyer}
\author[PURDUE]{Jungho Mun}
\author[INRIM]{Lorenzo Pattelli}
\author[Erlangen]{Lukas Pflug}
\author[KIT]{Carsten Rockstuhl\corref{cor1}}
\ead{carsten.rockstuhl@kit.edu}
\cortext[cor1]{Corresponding author}
\author[POSTECH]{Junsuk Rho}
\author[Wien]{Stefan Rotter}
\author[Marseille]{Brian Stout}
\author[Aalto]{P\"aivi T\"orm\"a}
\author[Laguna]{Jorge Olmos Trigo}
\author[KIT]{Frank Tristram}
\author[Thessaloniki]{Nikolaos L. Tsitsas}
\author[PaulPascal]{Renaud Vall\'ee}
\author[Lyon]{Kevin Vynck}
\author[Graz]{Thomas Weiss}
\author[Toulouse]{Peter Wiecha}
\author[Bremen]{Thomas Wriedt}
\author[NTUAthens]{Vassilios Yannopapas}
\author[Rouen]{Maxim A. Yurkin}
\author[NTUAthens]{Grigorios P. Zouros}

\affiliation[KIT]{organization={Karlsruhe Institute of Technology},
            addressline={Kaiserstrasse 12}, 
            city={Karlsruhe},
            postcode={76131}, 
            country={Germany}}

\affiliation[EPFL]{organization={\'Ecole Polytechnique F\'ed\'erale de Lausanne},
            addressline={Route Cantonale}, 
            city={Lausanne},
            postcode={1015}, 
            country={Switzerland}}

\affiliation[Aalto]{organization={Aalto University School of Science}, 
addressline={PO Box 15100}, 
            city={Aalto},
            postcode={00076}, 
            country={Finland}}

\affiliation[Wellington]{organization={The Macdiarmid Institute for Advanced Materials and Nanotechnology, School of Chemical and Physical Sciences, Victoria University of Wellington},
 addressline={PO Box 600}, 
            city={Wellington},
            postcode={6140}, 
            country={New Zealand}}
            
\affiliation[TUHH]{organization={Hamburg University of Technology},
 addressline={Ei\ss endorfer Stra\ss e 42}, 
            city={Hamburg},
            postcode={21073}, 
            country={Germany}}

\affiliation[PaulPascal]{organization={Univ. Bordeaux, CNRS, CRPP},
 addressline={UMR 5031}, 
            postcode={F-33600}, 
            city={Pessac},
            country={France}}

\affiliation[PTB]{organization={Physikalisch-Technische Bundesanstalt},
 addressline={Bundesallee 100}, 
            city={Braunschweig},
            postcode={38116}, 
            country={Germany}}

\affiliation[ZIB]{organization={Zuse Institute Berlin},
 addressline={Takustrasse 7}, 
            city={Berlin},
            postcode={14195}, 
            country={Germany}}

\affiliation[POSTECH]{organization={ Pohang University of Science and Technology (POSTECH)},
            city={Pohang},
            postcode={37673}, 
            country={Republic of Korea}}
            
\affiliation[UniWarsaw]{organization={Faculty of Physics,
University of Warsaw},            addressline={Pasteura 5}, 
            city={Warsaw},
            postcode={PL-02-093}, 
            country={Poland}}

\affiliation[Hannover]{organization={Leibniz University Hannover},
            addressline={Welfengarten 1}, 
            city={Hannover},
            postcode={30167}, 
            country={Germany}}

\affiliation[Wien]{organization={Vienna University of Technology (TU Wien)},
            addressline={Wiedner Hauptstra\ss e 8-10/136}, 
            city={Vienna},
            postcode={1040}, 
            country={Austria}}

\affiliation[Graz]{organization={Institute of Physics, University of Graz},
            addressline={Universit\"atsplatz 5}, 
            city={Graz},
            postcode={8010}, 
            country={Austria}}     

\affiliation[LP2N]{organization={Institut d'Optique d'Aquitaine},
            addressline={1 Rue F. Mitterrand}, 
            city={Talence},
            postcode={33400}, 
            country={France}}  

\affiliation[PURDUE]{organization={Purdue University},
            city={West Lafayette},
            postcode={IN 47904}, 
            country={USA}}  

\affiliation[INRIM]{organization={Istituto Nazionale di Ricerca Metrologica},
            addressline={Str. delle Cacce 91}, 
            city={Turin},
            postcode={10135}, 
            country={Italy}}  

\affiliation[Erlangen]{organization={Friedrich-Alexander-Universit\"at Erlangen-N\"urnberg},
            addressline={Martensstr. 5a}, 
            city={Erlangen},
            postcode={91058}, 
            country={Germany}} 

\affiliation[Thessaloniki]{organization={Aristotle University of Thessaloniki},
            city={Thessaloniki},
            postcode={54124}, 
            country={Greece}}     

\affiliation[Marseille]{organization={Universit\'e d'Aix-Marseille},
            addressline={52 Av. Escadrille Normandie Niemen}, 
            city={Marseille},
            postcode={13013}, 
            country={France}}                 

\affiliation[Laguna]{organization={Universidad de La Laguna},
            addressline={Apdo. 456}, 
            city={San Crist\'obal de La Laguna},
            postcode={38200}, 
            country={Spain}} 

\affiliation[Lyon]{organization={Institut Lumi\`ere Mati\`ere, CNRS, Universit\'e Claude Bernard Lyon 1},
            addressline={10 Rue Ada Byron}, 
            city={69622 Villeurbanne Cedex},
            country={France}} 

\affiliation[Toulouse]{organization=
{LAAS, Universit\'e de Toulouse, CNRS}, 
            city={Toulouse},
            country={France}} 

\affiliation[Bremen]{organization={Universit\"at Bremen},
            addressline={Badgasteiner Str.3}, 
            city={Bremen},
            postcode={28359}, 
            country={Germany}} 

\affiliation[NTUAthens]{organization={National Technical University of Athens},
            addressline={Iroon Polytechniou St. 9}, 
            city={Athens},
            postcode={15780}, 
            country={Greece}} 

\affiliation[Rouen]{organization={ Universit\'e Rouen Normandie, INSA Rouen Normandie, CNRS, CORIA UMR 6614},
            city={Rouen},
            postcode={F-76000}, 
            country={France} }

\begin{abstract}
The transition matrix, frequently abbreviated as T-matrix, contains the complete information in a linear approximation of how a spatially localized object scatters an incident field. The T-matrix is used to study the scattering response of an isolated object and describes the optical response of complex photonic materials made from ensembles of individual objects. T-matrices of certain common structures, potentially, have been repeatedly calculated all over the world again and again. This is not necessary and constitutes a major challenge for various reasons. First, the resources spent on their computation represent an unsustainable financial and ecological burden. Second, with the onset of machine learning, data is the gold of our era, and it should be freely available to everybody to address novel scientific challenges. Finally, the possibility of reproducing simulations could tremendously improve if the considered T-matrices could be shared. To address these challenges, we found it important to agree on a common data format for T-matrices and to enable their collection from different sources and distribution. This document aims to develop the specifications for storing T-matrices and associated metadata. The specifications should allow maximum freedom to accommodate as many use cases as possible without introducing any ambiguity in the stored data. The common format will assist in setting up a public database of T-matrices.
\end{abstract}

\end{frontmatter}


\part{Introduction}

\section{Background information}
The transition matrix, or T-matrix, constitutes a comprehensive representation of the optical properties of a scatterer in linear approximation \cite{waterman1965matrix,mishchenko2013peter}. In that context, the basic scattering problem can be expressed as follows: Given a scatterer in the surrounding medium and all its properties, i.e., shape and material composition, what is the scattered field for a given illumination? Due to the restriction to linear response, we can solve the problem in the frequency domain, i.e., we consider time-harmonic excitation. The response to a pulsed illumination can be reconstructed thanks to the superposition principle \cite{Nieminen2003}. The advantage of the T-matrix approach resides in the fact that an algebraic expression, i.e., a matrix-vector-product, describes the light-matter interaction. For that, the incident and the scattered fields are expanded in a basis set, and their amplitudes are stored in a vector \cite{bohren2008absorption}. The T-matrix multiplied by the incident field vector gives the scattered field vector. The number of expansion coefficients is truncated in numerical calculations,  which lends the T-matrix a finite dimension. For the expansion, vector spherical harmonics are usually used to reflect the three-dimensional localized character of the objects. The lowest-order expansion coefficients capture the dipolar, quadrupolar, and octupolar responses, and higher orders are of interest as well \cite{nevcada2021multiple,liu2017multipolar,mun2019importance}. It remains to be mentioned that other matrices represent the optical response from a scattering object as well. A typical example would be the S-matrix that relates incoming and outgoing fields instead of incident and scattered fields. Moreover, the K-matrix, also called the reaction matrix, exists. It relates the total field outside the scatterer as a superposition of regular fields and singular fields~\cite{colom2016optimal}. The reaction matrix has the nice property that it is Hermitian for lossless systems. Still, the different matrices can be converted to each other. We concentrate, therefore, on one of them here, the T-matrix.

Besides being the basis for discussing the scattering response from a given object, the T-matrix allows the study of advanced photonic materials made from a larger number of objects on semi-analytical grounds \cite{mishchenko1996t,mackowski1996calculation}. Examples are coupled particles, ensembles of thousands, or even millions of identical or different scatterers that form amorphous photonic materials~\cite{Pattelli:18}, or infinite arrangements of identical unit cells that contain one or multiple scatterers \cite{khlebtsov2013t,theobald2021simulation,mackowski2022extension}. Scattering interaction with a substrate can also be considered \cite{czajkowski2020multipole}. Additionally, a computed T-matrix allows us to study the scattering by aggregates of particles or orientation-averaged scattering \cite{ustimenko2021multipole,fazel2022orientation}. This is needed in optical particle characterization and in atmospheric radiative transfer. The spectral domain of interest and the application to be explored on the base of T-matrices are diverse. In electrodynamics, it spans multiple scientific disciplines, such as optics and photonics, nanotechnology, astronomy and astrophysics, remote sensing, atmospheric science, biophysics, and nanomedicine~\cite{vandenbroucke2020costuum, olmos2024capturing, Wriedt2006, tsang2022theory}, and it can even be used for tasks such as measuring the size of air bubbles in water~\cite{hansen1985mie}. Beyond science, the description of scattering of light by particles also covers many important applications in metrology and technologies such as nanoelectronics and advanced material characterizations. 

To perform all this research, the T-matrix of an object needs to be known. While it can be obtained experimentally~\cite{olmos2024solving}, we discuss other approaches in the course of the paper. Unfortunately, analytical solutions are only available for basic shapes. For spheres, Gustav Mie finalized the solution of the problem in 1908 and gave us what we call nowadays the Lorentz-Mie coefficients~\cite{mishchenko2009gustav}. They form the entries of a diagonal T-matrix. For all other particles, numerical methods are required to obtain their T-matrix. The null-field or extended boundary condition methods can compute T-matrices of gyroelectric spherical objects~\cite{almpanis2021spherical}, as well as of dielectric and gyrotropic non-spherical objects \cite{9395390, barber1975scattering,zouros2022scattering}. Otherwise, any available Maxwell solver can be employed to compute the scattered field or the induced current in the scatterer for a set of different illumination conditions \cite{demesy2018scattering}. From the induced response, the T-matrix can be constructed. In the most direct approach, an individual vector spherical harmonic is used for the illumination, and the scattered response is expanded to obtain one column of the T-matrix. But also, plane wave illuminations are possible.

However, these computations consume quite some resources. Depending on the problem details, many full-wave simulations are necessary. To quantify the efforts, we may argue that the T-matrix might be of interest in dipolar or octupolar order, leading to six or 30 expansion coefficients for the field \cite{somerville2015accurate}. Generally, we need $2N(N+2)$ expansion coefficients for a T-matrix of order $N$. Given that the size of the T-matrix is $2N(N+2)\times 2N(N+2)$, we need $2N(N+2)$ simulations to retrieve the T-matrix for an object with no symmetry. Suppose we are interested in a dispersive T-matrix (e.g., for 200 wavelengths), one might wish to vary one or multiple geometrical parameters (e.g., a helix can be parameterized with at least four numerical parameters, and we might be interested in testing ten values for each parameter), and assume a computational time of our full wave solver of three minutes for one full wave problem. In that case, we may need something between three hours and three years to compute all these T-matrices. The upper limit is certainly an extreme example. Still, experience says that we need a few hours of computational time on a reasonable infrastructure to retrieve a T-matrix of interest with the necessary precision. Because a larger community is interested in the optical response from the same objects, this calculation and the possible re-calculation constitute a major challenge for different reasons.              

\section{Challenges}

It is intellectually not satisfying to repeat the same tasks multiple times. Therefore, it seems wise to calculate T-matrices once, systematically store them, and make them available for later reuse according to the principles of findability, accessibility, interoperability, and reusability (FAIR principles) \cite{scheffler2022fair,ryabchykov2024photonic,heidenreich2023strategic}. Beyond this general consideration, more practical reasons strongly suggest a common data format for T-matrices and their proper storage.

First, their computation consumes resources, both intellectual and scientific, as it requires dedicated scientists to handle the computation. Still, conventional resources are equally needed. Besides hardware, the energy used to perform the calculations should not be underestimated. With the onset of the current energy crisis, we, as a community, are asked to perform resource efficient computations. But even before, the increased energy expenses of computing facilities put the issue of reducing energy consumption on the agenda. To quantify the importance, data from the Scientific Computing Center at the Karlsruhe Institute of Technology suggests that the computational power per investment into hardware doubles every 1.5 years. In contrast, the computational power per energy consumption doubles only every 2.7 years. Taking the ratio implies that the energy consumption per investment doubles every 3.3 years. That trend appears to be stable for the past ten years. As of today, the expenses for energy within five years correspond to the investment sum for a contemporary supercomputer. Following this trend of increasing importance of electricity costs suggests that in ten years, the expenses for energy will be eight times higher than for hardware. And in 20 years, the expenses for energy will be even 64 times higher. Consequently, financial support for energy, rather than hardware, will be the more crucial issue in the years to come.

Beyond that economic challenge, there is the associated environmental challenge. We shall strive to reduce our carbon footprint as much as possible to contribute to the solution of an essential problem for humanity, i.e., climate change. One answer could be to stop computations. But this does not make much sense when our purpose is to achieve progress and contribute to solving problems humanity faces. Photonic structures find use in devices for energy harvesting, photocatalysis, water purification, cancer treatment, and many more. We should not stop doing research. However, we should strive to perform research responsibly. Avoiding repetitive calculation of T-matrices on the base of which we study optical systems is one contribution that our community should aim at.

Moreover, performing science reproducibly is increasingly essential. Publishing computational codes open source is one step in this direction. Especially in the context of the T-matrix-based scattering formalism, we witnessed in the past years the publication of multiple codes for that purpose that all have their unique focus \cite{Doicu2006,Martin2019, nevcada2021multiple, beutel2024treams, egel2021smuthi, egel2017celes, schebarchov2022multiple,stefanou2000multem, somerville2016smarties}. With the publication of these codes, we did not just give back to the public what had been supported by taxpayers' money. We also put others in the position to reproduce our results, develop them further, or contribute in new directions. This empowers an entire community and generates trust in published results. 

However, a key ingredient in the multiple scattering formalism, the actual T-matrix, frequently cannot be generated within the framework of the semi-analytical scattering theory. Instead, it requires additional software that solves Maxwell's equations to generate T-matrices. This necessity of additional software might be a burden, and the cascading of research tools reduces the transparency of the numerical work. Relying on publicly available T-matrices would lower that dependency. 
In this context, the T-matrix is not only valuable as a subject of interest on its own, but serves as an interface between different computation tools.

Many computer codes have been developed over the years to compute T-matrices. Here, we focus on methods to compute scattering by non-axisymmetric scatterers. The T-matrix of a non-axisymmetric object (an ellipsoid) was first computed via the Null-field Method (NFM) (also referred to as the Extended Boundary Condition Method, EBCM) by Barber in his Ph.D. thesis \cite{Barber1973}. Schneider and Peden used the NFM to compute scattering by an ellipsoid \cite{Schneider1988}. Later, Laitinen, and Lumme \cite{Laitinen1998} used the method to compute scattering by a cube expanded into spherical functions. Wriedt \cite{Wriedt2002} and Doicu \textit{et al.} \cite{Doicu2006} used the NFM with discrete sources (NFM-DS) to compute scattering by arbitrarily shaped 3D particles described by a triangulated surface. Yurkin and Kahnert also computed scattering by cubes and compared results to that of the discrete dipole approximation (DDA)~\cite{Yurkin2013}. The T-matrix of a scattering object can also be computed via other surface and volume based computational electromagnetics methods. Nieminen \textit{et al.} \cite{Nieminen2003} used the point matching method. Das \textit{et al.} \cite{Das2023} used the surface integral equation method (SIEM). Mackowski \cite{Mackowski2002} as well as Loke \textit{et al.} \cite{Loke2009} used DDA. A similar method, the volume integral equation method (VIEM) is used by Markkanen and Yuffa \cite{Markkanen2017}. Recently, there is much research in the invariant embedding method to compute the T-matrix of complex shaped particles. This method is used by Bi \textit{et al.} \cite{Bi2013} and Doicu \textit{et al.} \cite{Doicu2019}. Generally, there is a larger number of general purpose solvers available to compute the T-matrix of a scatterer~\cite{scattport}.

Finally, we want to emphasize the positive aspects of having an agreed T-matrix data format and a database where T-matrices are collected and archived. Hosting the template scripts in a dedicated repository and including source files for each datafile contributes to the comparability and interchangeability of the programs and promotes collaborations in the community. With that, it supports the validation and the identification of possible systematic errors of the different approaches 

New research questions emphasizing data-intensive methods can be tackled based on aggregated T-matrices. For example, training neural networks that solve direct or inverse problems in scattering theory would become more feasible \cite{so2019simultaneous,talebi2020inferring}. This, in essence, could avoid one day even the necessity of an ordinary Maxwell solver to obtain the T-matrix of a given structure. Whether such networks can be trained for general purposes remains an open question, but this should be possible for important geometries frequently considered. Also, some basic science questions could be answered, like whether there is a particle for any possible T-matrix that copes with all physical constraints that can be imposed. Collecting these T-matrices stored in an agreed data format would be a stepping stone for entirely new directions our communities could pursue.

For all these reasons, the following document contains information on a data format we suggest using in the future. The document has been worked out by a larger number of groups working on the development of tools to study scattering problems. That ranges from atmospheric scattering problems, to nanophotonics, to metrology, to scattering in biological samples and beyond. It is the effort of a larger number of community members \cite{antosiewicz2014plasmonic, agocs2015scatterometry}.  

The document includes two parts after a brief introduction on the T-matrix formalism in the following. First, the data format specifications are described. Afterward, multiple aspects concerning the generation and validation of the T-matrices are discussed. Then, we outline some details related to the exploitation of the generated files in multiscattering problems and provide some summarizing statements in the end. Along with this documentation, codes that contain utilities to work with the data format are provided under the following link:  \\
\href{https://github.com/tfp-photonics/tmatrix_data_format}{https://github.com/tfp-photonics/tmatrix\_data\_format}
\,.\\ In perspective, a data repository and a dedicated web interface will be made available as a preferred portal for the archiving and exchanging of T-matrix datasets. However, this description would exceed the scope of the current manuscript, and we leave its documentation for a future article. 

\section{T-matrix formalism}

\begin{figure}[h!!]
  \centering
  \includegraphics[width=0.8\linewidth]{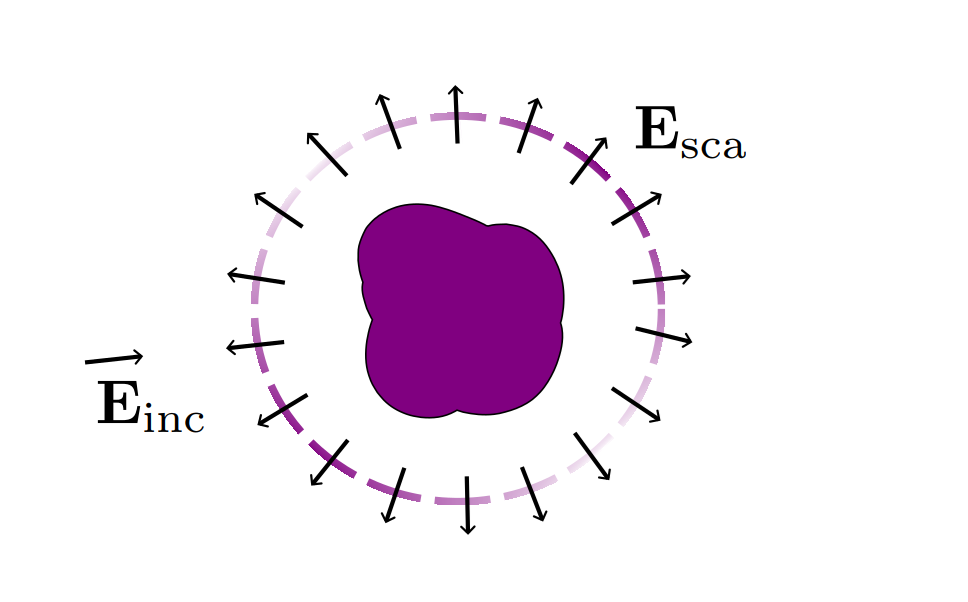}
  \caption{Schematic representation of the scattering problem for an arbitrary scatterer in free space.}
  \label{scatproblem}
\end{figure}

Here, we support the concepts introduced above with mathematical formulations. The discussion is kept rather brief and only given with the purpose to set a common ground to discuss the data format. A recent general introduction into the method can be found in the book chapter by Mackowski~\cite{MACKOWSKI2023113}. 
The scattering problem is illustrated in Fig.~\ref{scatproblem}.
It is expressed with the following equation:
\begin{equation}
    \mathbf{p} =  \mathbf{T}  \mathbf{a}\, ,
\end{equation}
where the vectors $\mathbf{p}$ and $\mathbf{a}$ represent the expansion coefficients of scattered and incident fields, respectively, and $\mathbf{T}$ is the T-matrix \cite{alaee2019exact}. The expansion of the scattered field is not considered to be valid for nonspherical scatterers everywhere inside the smallest sphere circumscribing the scatterer, which is a topic of ongoing research~\cite{schebarchov2019mind, lamprianidis2023transcending, bertrand2020global}. The expansion can be performed with different basis functions, the most common one being the vector spherical wave functions (VSWF): 

\begin{align}
  \mathbf{E}_{\text{inc}}(\mathbf{r}, \omega) &= \sum_{l=1}^{\infty}\sum_{m=-l}^{l}\left[a_{lm}^\mathrm{e}(\omega) \mathbf{N}^{(1)}_{l m }(\mathbf{r}, \omega) + a_{lm}^\mathrm{m}(\omega) \mathbf{M}^{(1)}_{l m }(\mathbf{r}, \omega)\right]
  \\
    \mathbf{E}_{\text{sca}}(\mathbf{r}, \omega) &= \sum_{l=1}^{\infty}\sum_{m=-l}^{l}\left[ p_{lm}^\mathrm{e} (\omega) \mathbf{N}^{(3)}_{l m }(\mathbf{r}, \omega) + p_{lm}^\mathrm{m}(\omega) \mathbf{M}^{(3)}_{l m }(\mathbf{r}, \omega)\right]\, .
\end{align}
In these expressions, VSWFs are defined in parity basis, $\mathbf{M}_{l m }(\mathbf{r}, \omega)$ are the transverse electric (TE) or magnetic multipole fields, and $\mathbf{N}_{l m }(\mathbf{r}, \omega)$  are the transverse magnetic (TM) solutions or electric multipole fields~\cite{jackson1998classical}. Beware that different authors use different definitions of spherical vector wave
functions. The superscripts (1) and (3) correspond to the choice of the spherical Bessel or Hankel functions in the solution, resulting in regular or singular solutions.
The field inside the scatterer can also be expanded in VSWFs:
\begin{equation}
  \mathbf{E}_{\text{int}}(\mathbf{r}, \omega) = \sum_{l=1}^{\infty}\sum_{m=-l}^{l}\left[c_{lm}^\mathrm{e}(\omega) \mathbf{N}^{(1)}_{l m }(\mathbf{r}, \omega) + d_{lm}^\mathrm{m}(\omega) \mathbf{M}^{(1)}_{l m }(\mathbf{r}, \omega)\right]\, .
 \end{equation}
In the numerical calculation, the multipole order $l$ is truncated to the largest order non-negligible for a given problem.  A basis with well-defined helicity can be alternatively employed. 
The center of expansion can also be chosen in different ways~\cite{kildishev2023art}.
 Alternatively, a cylindrical wave expansion is beneficial for problems with a specific geometry and can be used to analytically solve the scattering problem of an infinite cylinder with arbitrary geometrical cross-section~\cite{bohren2008absorption, katsinos2020entire}. However, this basis set will not be addressed in the further discussion. The exact definitions and the normalization used in the data format specification are noted in \ref{app:normalization}.

The VSWFs are complex-valued in the general case. Importantly, the time evolution is defined here by the factor $\exp(-\mathrm \mathrm{i} \omega t)$, which is more common in the optics community than its complex-conjugated counterpart. If required for some approaches, such as quasi-normal modes, complex frequencies can be considered~\cite{lalanne2018light,lalanne2019quasinormal}. Occasionally, the notion of \enquote{mode} here is used to refer to the vector spherical harmonics because they are simple solutions of the source-free Maxwell equations in homogeneous space in a spherical coordinate system, comparable to plane waves which are simple solutions to the Maxwell equations in homogeneous space in a Cartesian coordinate system. This is not to be confused with quasi-normal modes, which are the modes sustained by a scatterer.

An additional aspect worth mentioning is the connection of the T-matrix to the S-matrix, represented in a simple relation:
 \begin{equation}
     \mathbf{S} = \mathbb{1} + 2  \mathbf{T}\, ,  
 \end{equation}
where $\mathbb{1}$ is the identity matrix. This relation follows from the fact that the S-matrix connects incoming with outgoing modes (in contrast to incident and scattered modes). While normally the incident fields in the T-matrix formalism are expanded in spherical Bessel functions (regular) and scattered fields in spherical Hankel functions (singular), the fields in S-matrix formalism are normally expanded using spherical Hankel functions, thus the additional factor of two coming from the relation between them. Please note, alternative formulations for the relation can be found, e.g., $\mathbf{S} = \mathbb{1} + \mathbf{T}$~\cite{vavilin2024polychromatic}. But the validity of such expression requires some changes in the normalization of the basis function sets. The S-matrix is widely used with other basis functions as well, for example, eigenmodes of a waveguide~\cite{rotter2017light} or of an optical fiber~\cite{Cao:23}, and in the context of time-varying scattering processes~\cite{globosits2024photonic}. Another example for an alternative matrix expressing the scattering problem is the previously mentioned reaction matrix, also called K-matrix, which can be obtained via $\mathbf{K}=\mathrm{i}\mathbf{T}\left(\mathbb{1}+\mathbf{T}\right)^{-1}$ from the T-matrix~\cite{colom2016optimal}.

Building upon these concepts, the T-matrix of an arrangement of scatterers can be formulated with the help of the translational addition theorem for VSWFs~\cite{cruzan1962translational}.     
The following equation can be derived: 
\begin{equation}\label{toglobal}
 \mathbf{p}_{\text{local}}= (\mathbb{1} - \mathbf{T}_{\text{local}} \mathbf{C}^{(3)}_{\text{local}})^{-1} \mathbf{T}_{\text{local}} \mathbf{a}_{\text{local}}
\, , 
\end{equation}
where \enquote{local} in the index implies that the T-matrices of all the objects comprising the total T-matrix are defined in their local coordinate systems, and $\mathbf{C}_{\text{local}}^{(3)}$ contains the translation coefficients. Moreover, other formulations to solve the same problem exist, and they all can be employed on the base of the T-matrices~\cite{stout2008recursive}. Finally, for many scatterers, it can be computationally more efficient to shift all the matrices to a common origin and diminish their size by doing so \cite{dwivedi2024effective}. This is achieved by multiplying translation coefficients with the expansion coefficients as well:

\begin{equation}
\mathbf{p}_{\text{global}}=\left( \mathbf{C}_{01}^{(1)} ... \mathbf{C}_{0N}^{(1)} \right) \mathbf{p}_{\text{local}}
\end{equation}
and
\begin{equation}
\mathbf{a}_{\text{local}}= \begin{pmatrix}\mathbf{C}_{10}^{(1)} \\ \vdots \\ \mathbf{C}_{N0}^{(1)} \end{pmatrix} \mathbf{a}_{\text{global}}\, ,
\end{equation}
such that the global T-matrix connects the two newly defined vectors~\cite{suryadharma2017studying}. 

With the T-matrix at hand, useful characteristics of the scattering response can be derived directly~\cite{mishchenko1996t}, such as the orientation-averaged extinction cross-section:
\begin{equation}
\langle \sigma_{\text{ext}}\rangle = 
-\frac{2\pi} {k^2} \mathrm{Re} \sum_{l=1}^{l_{\text{max}}} \sum_{m = -l}^{l} \sum_{i= \pm 1} T^{ii}_{l,m,l,m}\, ,\end{equation}
and the orientation-averaged scattering cross-section:
\begin{equation}
\langle \sigma_{\text{sca}}\rangle = 
\frac{2\pi} {k^2} \sum_{l=1}^{l_{\text{max}}}  \sum_{l'=1}^{l_{\text{max}}} \sum_{m = -l}^{l} \sum_{m' = -l'}^{l'}  \sum_{i=\pm 1}  \sum_{j= \pm 1} \lvert T^{ij}_{l,m,l',m'} \rvert ^2\, ,\end{equation}
where $k$ is the wavenumber, $l \in \mathbb N_0$, $m \in \mathbb Z$ with $l \geq |m|$, and $i, j$  indices indicate the polarization of the incident and scattered modes, respectively, and take the values of -1 and 1. These expressions were derived assuming illumination with plane waves. In the general case, the cross-sections are only proportional to the right-hand side of the equations. The cross-sections are always positive-defined. 

The interaction between scatterers in an infinite periodic arrangement is also possible in this framework with the help of the Ewald summation technique~\cite{linton2010lattice}, enabling the computation of metasurface parameters of interest, such as transmittance and reflectance \cite{schulz2024roadmap}. The concise representation of the optical response is particularly advantageous for more complex geometries of interest that include scatterers with known T-matrices. It facilitates a considerably faster treatment than the calculation of scattering response for each particular arrangement and optical characteristic from scratch using a full-wave Maxwell's solver \cite{yannopapas2007negative,achouri2022multipolar}. 

From all these discussions, we see that the T-matrix is central for the exploration of the optical properties of scattering systems. Therefore, we strive to develop a data format in the following that allows to systematically create, store, and share it. 

\part{Data format specifications} \label{Data format specification}
\setcounter{section}{2}
This part describes the specification of the data format for storing T-matrices. It is intended to give all the necessary details to generate valid files containing T-matrix data and to develop the necessary tools. We stress that many different formats could have been chosen. Still, after intense discussions among the expert colleagues who authored the document at hand over an extended period of time and a dedicated workshop that took place in Bad Herrenalb, Germany in December 2023, an agreement was found on a suitable set of specifications. The following details are motivated by some basic requirements that we impose on the storage format. In particular, it needs to include:
\begin{enumerate}
    \item Clear definitions of the T-matrix, especially regarding the vector spherical waves and their normalization,
    \item Unique descriptions of the properties of a scatterer,
    \item Comprehensive descriptions of the computation method for reproducibility, and
    \item An accessible storage format to support different software.
\end{enumerate}
Each file can have optional components along with the required components specified. 
We describe the different components of the data file, assuming a hierarchical data structure in the HDF5 format. We provide a brief overview of that format in \ref{sec:hdf5}, as well as an introduction of tools to convert to this standard from some existing formats.  It is recommended to store the files with the extension \texttt{.tmat.h5} (or \texttt{.tmat.hdf5}) to highlight the specific structure of the file. In the file naming scheme, we recommend rounding the numerical values to the second significant digit after the comma. In the following, we demonstrate an example of the file structure for visual assistance. The font is selected to emphasize the usage of \texttt{\textbf{groups}}, \texttt{datasets}, and \texttt{\textit{attributes}}.

\dirtree{%
.1 hdf5 file.
.2 \textit{name}, \textit{description}, \textit{keywords}, \textit{storage\_format\_version}.
.2 tmatrix.
.2 \bf{modes}.
.3 l.
.3 m.
.3 polarization.
.2 frequency.
.3 \textit{unit}.
.2 \bf{embedding}.
.3 \textit{name}, \textit{description}, \textit{keywords} .
.3 relative\_permittivity.
.3 relative\_permeability.
.2 \bf{scatterer}.
.3 \bf{material}.
.4 \textit{name}, \textit{description}, \textit{keywords} .
.4 relative\_permittivity.
.4 relative\_permeability.
.3 \bf{geometry}.
.4 \textit{name, description, keywords, shape, unit}.
.4 \textit{...}.
.2 \bf{computation}. 
.3 \textit{software},  \textit{method}, \textit{name}, \textit{description}, \textit{keywords}. 
.3 \bf{files}.
.3 mesh.XYZ.
.3 \bf{method\_parameters}.
}
\hspace{0.5cm}
The required entries given in the following subsections should not be omitted, unless it is mentioned explicitly, since they provide important information concerning the T-matrix and the scatterer of interest. 

\subsection{T-matrix}

The \colorbox{mygray}{\lstinline|/tmatrix|} is the main \emph{dataset} of the file. It contains an array of complex values with the shape $(a, b_s, b_i)$. The last two dimensions are the $b_s \times b_i$ entries of the T-matrices with $b_s$ the number of scattered modes and $b_i$ the number of incident modes. The wavelength-related sweep of length $a$ is currently the only allowed third dimension for each T-matrix. Within an individual file, no other parameters describing the computation and the scatterer should be varied, unless it concerns the dispersive nature of material parameters. This is for the sake of simplicity of reading and searching the metadata. 
The last two dimensions should be related to the modes in the \emph{group} \colorbox{mygray}{\lstinline|/modes|}.
In other words, the \emph{attribute} {\colorbox{mygray}{\lstinline|inner_dims|}} is set implicitly to 2. Any other explicit value is prohibited.

An optional dataset to store is the T-matrix connecting the incident fields and the internal fields.  It can be saved as a separate dataset \colorbox{mygray}{\lstinline|/rmatrix|}. This matrix should be associated with the same modes as the main T-matrix and have the same normalization, with all the restriction in the previous paragraph applicable to its dimensions. The background medium used in the computation of this R-matrix is assumed to be the material of the scatterer.
\subsection{Name, description, and keywords}

Each T-matrix has a string \colorbox{mygray}{\lstinline|/name|} that
concisely describes the T-matrix and possibly its distinguishing feature. It is stored as an \emph{attribute} of \colorbox{mygray}{\lstinline|/|}.
While a unique notation or wording for \colorbox{mygray}{\lstinline|/name|} is not required, this attribute helps in the identification and retrieval of the T-matrix as well as for its usage for machine learning and other methods of analysis. Also, a unique identifier will be assigned automatically during the upload of data to the database, which will be discussed elsewhere.

The \emph{attribute} \colorbox{mygray}{\lstinline|/description|} is a string that comprehensively outlines the object, its shape, and materials. It can also contain additional information, e.g., optimization goals that were attempted to be reached with the specific geometry. If the data was computed with a particular application in mind, the application field can be separated into optional \emph{attribute} \colorbox{mygray}{\lstinline|/application|} to simplify search in the future database. After reading the name and description, one should have a precise idea of what the T-matrix describes. A comma-separated list of important keywords can be added to the file as well as the \emph{attribute} \colorbox{mygray}{\lstinline|/keywords|}. These three attributes are generally useful to be added in the materials, computation, or geometry definitions. The keywords can be used to provide information on special properties such as symmetries. It is helpful to follow consistent naming for the symmetries, the common ones being \textit{czinfinity} (rotational symmetry), \textit{mirrorxyz} and all possible combinations of symmetry planes, \textit{reciprocal}, \textit{passive}, \textit{lossless}. The details of their definitions are provided in~\Cref{symmetries}.  We stress that carefully chosen names and an elaborative description will help at a later stage to retrieve and reuse a given entry of the dataset. To ensure optimal compatibility with machine learning frameworks, adherence to a standardized keyword and property scheme in the metadata can be supported by using machine learning methods such as large language models to identify unifying/reoccurring elements in the metadata.



\subsection{(Angular) frequency, vacuum wavelength, or (angular) wavenumber}

The frequency of the electromagnetic waves needs to be known to understand the T-matrix. In different communities, different ways of describing this information are common. For machine learning applications, it is essential that the frequency-related data are consistently formatted and includes clear units. This ensures that models can accurately interpret and utilize the data for training and prediction tasks. It is allowed to submit not only real values, but also complex values for methods that exploit quasi-normal modes~\cite{wu2020intrinsic}. Adding multiple descriptions, e.g., frequency and wavelength data, is discouraged. The relation between the datasets is defined by
\[
    \frac{2 \pi \nu}{c_0}
    = \frac{\omega}{c_0}
    = \frac{2 \pi}{\lambda_0}
    = 2 \pi \tilde{\nu_0} = k_0
\]
with the speed of light in vacuum $c_0 = \SI{299792458}{\metre\per\second}$.
\begin{table}
    \centering
    \caption{
        Different ways to define the frequency. \label{tab:freqlambda}
    }
    \begin{tabular}{llc}
        \toprule
        Dataset name & Name & symbol \\
        \midrule
        \colorbox{mygray}{\lstinline|/frequency|} & Frequency & $\nu$ \\
        \colorbox{mygray}{\lstinline|/angular_frequency|} & Angular frequency
            & $\omega$ \\
        \colorbox{mygray}{\lstinline|/vacuum_wavelength|} & Vacuum wavelength
            & $\lambda$ \\
        \colorbox{mygray}{\lstinline|/vacuum_wavenumber|} & Vacuum wavenumber
            & $\tilde{\nu}$ \\
        \colorbox{mygray}{\lstinline|/angular_vacuum_wavenumber|}
            & Angular vacuum wavenumber & $k$ \\
        \bottomrule
    \end{tabular}
\end{table}
The \emph{dataset} has the required \emph{attribute} \colorbox{mygray}{\lstinline|unit|}, that defines the SI unit and prefix of the data as a string, e.g., \enquote{\si{\tera\hertz}} for the (angular) frequency or \enquote{\si{\centi\meter}\textasciicircum\{\nobreakdash-1\}} for the (angular) wavenumber. Frequencies can also be expressed in inverse seconds, for example, \enquote{\si{\second}\textasciicircum\{\nobreakdash-1\}}. The unit micrometers can be expressed as \enquote{\si{\micro\meter}} or \enquote{u\si{\meter}}.
There is no unit assumed by default. The full list of accepted units is given in \ref{app:units}.  
\subsection{Modes}
Since there are many ways to sort the entries of the T-matrix, the related mode of each row and column has to be explicitly given. This data is collected in the \emph{group} \colorbox{mygray}{\lstinline|/modes|}. The definition of the modes is given in \ref{app:normalization}. These modes are indexed by a degree $l \in \mathbb{N}$ and by an order $m \in \mathbb{Z}$ with $|m| \leq l$. This data is given in the corresponding \emph{datasets} \colorbox{mygray}{\lstinline|/modes/l|} and \colorbox{mygray}{\lstinline|/modes/m|}. Additionally, the polarization is given in \colorbox{mygray}{\lstinline|/modes/polarization|}. To avoid any ambiguity, the polarization of each mode is given as a string \enquote{electric} (also known as \enquote{TM}, for modes $\mathbf{N}_{lm}^{(n)}$), followed by \enquote{magnetic}  (\enquote{TE}, for modes $\mathbf{M}_{lm}^{(n)}$). If the helicity basis is used, the polarization is defined by alternating \enquote{positive} and \enquote{negative} for $\mathbf{A}_{lm+}^{(n)}$ and $\mathbf{A}_{lm-}^{(n)}$, respectively.   
The incident and scattered modes can be separated into \colorbox{mygray}{\lstinline|/modes/l_incident|} and \colorbox{mygray}{\lstinline|/modes/l_scattered|} (and likewise for the other items in the \emph{group}). When present, they take precedence. This splitting is required if a different number of incident and scattered modes are used.

One must store the modes in a fixed order, and for visual assistance, a table can be found in \ref{app:reference}. The parameters sweep in the following sequence: the degree $l$ ranges from 1 to $l_{\text{max}}$, for each fixed $l$ there are modes with order $m$ traversing from $-l$ to $l$, and, lastly, for each set of $(l, m)$ there are two alternating modes \enquote{electric} and \enquote{magnetic} (or \enquote{positive} and \enquote{negative}). As outlined above, the term \enquote{electric} multipole field is not to be confused with \enquote{transverse electric} field.
The modes are required to be uniform for one file. Thus, they are -- unless they are scalar -- always given as one-dimensional arrays of length $b_s$ and $b_i$.

Finally, for the special case of a cluster of scatterers when the T-matrix is defined in the local basis of each scatterer, the \emph{datasets} \colorbox{mygray}{\lstinline|/modes/positions|} and \colorbox{mygray}{\lstinline|/modes/index|} should be added to provide the correspondence between the T-matrix entries and the local coordinate systems.  The index of incident and scattered modes can also be separated into \colorbox{mygray}{\lstinline|/modes/index_incident|} and \colorbox{mygray}{\lstinline|/modes/index_scattered|}. 
\subsection{Storage format version}


It is required to specify the version of the storage format in the 
\emph{attribute} \colorbox{mygray}{\lstinline|/storage_format_version|} as a string. With this document, we fix the standard format at \enquote{v1}, which is also reflected as the first two characters of the version specification of the T-matrix data format repository~\cite{data_format} in \colorbox{mygray}{\lstinline|software|} \emph{attribute} of \colorbox{mygray}{\lstinline|/computation|}.

%



\subsection{Scatterer}
The T-matrix can be computed for a cluster of scatterers. Information on each scatterer is then collected in separate groups. Each group comprises the  information on both geometry and material. The names of these groups are distinguished by assigning a number to each name, e.g., \colorbox{mygray}{\lstinline|/scatterer_X|}. For a single scatterer, \colorbox{mygray}{\lstinline|/scatterer|} is accepted as the name of the group. It will be further referred to as \colorbox{mygray}{\lstinline|/NAME|}. 
\subsubsection{Material}
The material is a \emph{subgroup} of the \emph{group} \colorbox{mygray}{\lstinline|/NAME|} describing the scatterer. In case of multiple scatterers, each \colorbox{mygray}{\lstinline|material|} \emph{subgroup} is inside the corresponding scatterer \emph{group}. The surrounding medium is not included in the scatterer \emph{group}. 
The description of the material is typically done by providing the permittivity and permeability or by defining the refractive index and the impedance. 
Each material \emph{group} can be annotated with a \colorbox{mygray}{\lstinline|name|}, \colorbox{mygray}{\lstinline|description|}, and \colorbox{mygray}{\lstinline|keywords|} as \emph{attribute}. The recommended way to add the \colorbox{mygray}{\lstinline|name|} \emph{attribute} of the material is to specify both the chemical formula and the common name if available (e.g., \enquote{Au, Gold}). The definitions here assume linear, homogeneous materials. The specific case of a layered structure can be accommodated by entering the material parameters as arrays, and separating the attributes corresponding to each layer with a semicolon in the string. The materials can be dispersive.  In each scatterer group, the wavelength dependence should occur along the first axis of the datasets describing the material parameters. The keywords can also contain the type of material, e.g., dielectric. Special information, such as nonlocality, can be added as a \colorbox{mygray}{\lstinline|keyword|}, and the details can be included in the \colorbox{mygray}{\lstinline|description|}. Then, the local permittivity contribution is included into the standard datasets introduced below for material parameters. If the values for the relative permittivity and permeability were used from an external source, this can be specified in the \colorbox{mygray}{\lstinline|reference|} \emph{attribute}, in the form of a reference paper or other link as a string. Alternatively, if the values were measured experimentally, the original datasets should be added under the \emph{subgroup} \colorbox{mygray}{\lstinline|/NAME/material/experimental_data|}, which includes the material parameters as described below and the corresponding frequencies/wavelengths with specified units. The method used for interpolation should be added as well in a string format as an \emph{attribute} \colorbox{mygray}{\lstinline|interpolation|}.
We separate cases with isotropic and anisotropic media, as well as cases with and without magnetoelectric coupling. This leads to the four classes of isotropic, biisotropic, anisotropic, and bianisotropic materials, that we describe below (see also \ref{app:constitutiverelations}).

\paragraph{Isotropic materials}
The material parameters can either be a relative permittivity and permeability or a refractive index and relative impedance. Either of the pairs are to be specified fully, and the values are not set by default to vacuum. If any of the former two parameters are present, all occurrences of refractive index and relative impedance will be ignored. The relative permittivity $\epsilon$ is given in 
\colorbox{mygray}{\lstinline|/NAME/material/relative_permittivity|} and the relative permeability $\mu$ is given in \colorbox{mygray}{\lstinline|/NAME/material/relative_permeability|}. 
Alternatively, neither permittivity nor permeability is defined, and the refractive index $n = \sqrt{\epsilon \mu}$ in \colorbox{mygray}{\lstinline|/NAME/material/refractive_index|} and the relative impedance $Z = \sqrt{\frac{\mu}{\epsilon}}$ in \colorbox{mygray}{\lstinline|/NAME/material/relative_impedance|} are defined.

\paragraph{Anisotropic materials}

For an anisotropic material, the above-mentioned \emph{datasets} can be used, except the relative impedance. If any of the \emph{datasets} is used for an anisotropic material, the \emph{attribute} \colorbox{mygray}{\lstinline|inner_dims|} has to be defined. Otherwise, by default, {\lstinline|inner_dims|} is assumed to be 0. If it is set to 1, the values of the last dimension are taken as the diagonal values of the 3-by-3 tensor. If it is set to 2, the \emph{dataset} contains the full tensor. Additionally, the \emph{attribute} \colorbox{mygray}{\lstinline|coordinate_system|} can be set to either \enquote{Cartesian}, \enquote{cylindrical}, or \enquote{spherical} (see \ref{app:coordinates}) to specify the chosen set of coordinates. The default is \enquote{Cartesian}. In the spherical or cylindrical case, the z-axis takes a special role by default (axis of rotation of axis-symmetric objects). This default can be changed by appending \enquote{x} or \enquote{y} to the \emph{attribute}, e.g., \enquote{cylindricalx}.

\paragraph{Biisotropic materials}
A biisotropic material has up to two additional parameters as an isotropic material. One of these two additional parameters is \colorbox{mygray}{\lstinline|/NAME/material/chirality|}. Moreover, a second additional parameter is \colorbox{mygray}{\lstinline|/NAME/material/nonreciprocity|}. Both parameters have the default value~0. Biisotropic parameters can only be defined in conjunction with relative permittivity and relative permeability. Using the refractive index or relative impedance is prohibited in that case.

\paragraph{Bianisotropic materials}

A bianisotropic material can be considered by using the parameters of the biisotropic materials in conjunction with the \colorbox{mygray}{\lstinline|inner_dims|} \emph{attribute}. Alternatively, it can be defined by giving the 6-by-6 bianisotropic tensor in the \emph{dataset}
\colorbox{mygray}{\lstinline|/NAME/material/bianisotropy|}. Other material parameters are not permitted in that case. The \emph{attribute} \colorbox{mygray}{\lstinline|inner_dims|} has to be set to 1 or 2 to specify if the bianisotropy is given as full tensor or as diagonal values only. (With \colorbox{mygray}{\lstinline|inner_dims|} set to 1 the material is, in fact, not bianisotropic but anisotropic.)

\subsubsection{Geometry}

The geometry of the objects described by the T-matrix can be defined as a  \emph{subgroup} \colorbox{mygray}{\lstinline|/geometry|} of the \emph{group} \colorbox{mygray}{\lstinline|/NAME|}. 
There are various ways to describe the geometry. Therefore, this section can be adapted. 
Again, a \colorbox{mygray}{\lstinline|name|}, \colorbox{mygray}{\lstinline|description|}, and \colorbox{mygray}{\lstinline|keywords|} can be added as \emph{attributes} for this \emph{group}. For an arrangement of scatterers, the \emph{dataset} \colorbox{mygray}{\lstinline|position|} is to be indicated explicitly for each scatterer. Position at the center of the coordinate system is assumed by default. The coordinates defined in \colorbox{mygray}{\lstinline|position|} specifies the center of the smallest circumscribing sphere of the scatterer. 

If the scatterer has a simple geometric shape, the parameters from Table 2 should be used to describe it. The shape should then be specified by adding the \emph{attribute} \colorbox{mygray}{\lstinline|shape|}. Note that by default for the rotationally symmetric objects, the symmetry axis is the z-axis. In the table,  core-shell sphere is separated into a separate entry, but for a general case of a layered scatterer, it is admissible to set the final shape of the scatterer as the \colorbox{mygray}{\lstinline|shape|} \emph{attribute} and add an array of geometrical parameters together with an array of material properties corresponding to each layer. The accepted convention is to measure the radius of the shell from the center of the whole object \cite{tsitsas2006scattering,langevin2024pymoosh}. If the same unit can be used to describe all geometrical parameters, it can be specified as an \emph{attribute} \colorbox{mygray}{\lstinline|unit|} of the \emph{group}. The unit can also be an \emph{attribute} of each individual parameter.

Clearly, the provided list of basic shapes cannot cover all scatterers that may be of interest, which limits the search capabilities within the future database. To address this, we strongly emphasize the importance of including the mesh in the data file, as detailed later in the section. The previously mentioned \colorbox{mygray}{\lstinline|name|} attribute can be considered as a possible entry to specify names for shapes not included in the basic shapes list. Upon reasonable request, new shapes will be added to the current list, which will be published in the GitHub repository.

The \emph{dataset} \colorbox{mygray}{\lstinline|/NAME/geometry/expansion_center|} is expected to specify the center of expansion of VSWFs used in the computation of the T-matrix, and by default, it is assumed to be at the center of the coordinate system.  For the listed objects with basic shapes, the coordinate system is fixed. For any rotated object with a basic shape, its final orientation can be specified in \colorbox{mygray}{\lstinline|/NAME/geometry/euler_angles|} \emph{dataset} using the standard Euler angles, following the extrinsic z-y-z notation. By default, these values are assumed to be zeros. For a freeform object, the orientation can only be inspected from the mesh file, since the standard orientation is not defined in the general case.


\begin{table}[t] 
\begin{minipage}{\textwidth} 
    \caption{Basic three-dimensional shapes and their defining parameters. }
    \centering \label{tab:shapes}  
    \begin{tabular}{|lp{85mm}|}
        \toprule
        Shape & Parameters \\
        \midrule
        \lstinline|sphere| & \lstinline|radius|  \\
        \lstinline|cut_sphere| \cite{Doicu2006} & \lstinline|radius|,  \lstinline|height|  \\
        \lstinline|core_shell_sphere|  &  \lstinline|radius_0| , \lstinline|radius_N|\\
        \lstinline|spheroid| & \lstinline|radiusxy|, \lstinline|radiusz|  \\
        \lstinline|ellipsoid|  & \lstinline|radiusx|, \lstinline|radiusy|, \lstinline|radiusz|  \\
        \lstinline|superellipsoid| \cite{Wriedt2002}  &  \lstinline|radiusx|, \lstinline|radiusy|, \lstinline|radiusz|, \lstinline|n_parm|, \lstinline|e_parm|   \\

        \lstinline|cylinder| \tablefootnote{~Rotational axis is along z by default for rotationally symmetric scatterers. For different orientations, please specify the Euler angles additionally as described in the main text.} & \lstinline|radius|, \lstinline|height| \\
        \lstinline|cone| & \lstinline|radius_top|,  \lstinline|radius_bottom|, \lstinline|height|
     \\
             \lstinline|ring| & \lstinline|radius_major|, \lstinline|radius_minor|, \lstinline|height|
           \\
    \lstinline|torus| & \lstinline|radius_major|, \lstinline|radius_minor|
           \\
   
        \lstinline|cube| & \lstinline|length|  \\
        \lstinline|rectangular_cuboid| & \lstinline|lengthx|, \lstinline|lengthy|,
            \lstinline|lengthz|  \\
       \lstinline|helix| \tablefootnote{~The pitch is oriented along the z-axis by default. For different orientations, please specify the Euler angles additionally as described in the main text.} \tablefootnote{~Since various definitions are possible, we stress here that the normal plane cross-section is the default. If this is not the case, please specify in the~\lstinline|description| \emph{attribute} of the geometry.} \tablefootnote{~\lstinline|termination| can be \enquote{spherical} or \enquote{flat}, \lstinline|handedness| can be \enquote{right} or \enquote{left}. The defaults are \enquote{flat} and \enquote{right}}.&  \lstinline|radius_helix|, \lstinline|radius_wire|,  \lstinline|pitch|, \lstinline|number_turns|, \newline
       \lstinline|termination|, 
       \lstinline|handedness|
       \\
     \lstinline|pyramid| &  \lstinline|n_edges|, \lstinline|radius|, \lstinline|height|, \lstinline|angle|,
      \lstinline|apex_shift|
     \\
     \lstinline|regular_prism| & \lstinline|n_edges|, \lstinline|radius|, \lstinline|height|, 
      \lstinline|shift|
     \\
     \lstinline|wedge| & \lstinline|lengthx|, \lstinline|lengthy|,
            \lstinline|lengthz|,  \lstinline|deltax|,  \lstinline|deltay| \\
 \lstinline|convex_polyhedron| & \lstinline|points|
     \\
        \bottomrule
    \end{tabular}
\end{minipage}
\end{table}

For a general representation of the scatterer, a mesh file has to be defined. The only possible reason for a mesh to be completely absent in the file is the choice of a semi-analytical method for computations (e.g., EBCM, Mie). In this case, \emph{semi-analytical} should appear in  \colorbox{mygray}{\lstinline|keywords|} \emph{attribute} of \colorbox{mygray}{\lstinline|/computation|}, and the presence of the \emph{attribute} \colorbox{mygray}{\lstinline|shape|} and the corresponding \emph{datasets} of geometrical parameters is obligatory. The mesh can be specified in \colorbox{mygray}{\lstinline|/NAME/geometry/mesh.XYZ|} or  \colorbox{mygray}{\lstinline|/computation/mesh.XYZ|}, where XYZ stands for the file extension, depending on what is physically more reasonable. For example, for multiple scatterers and a single mesh file, it is sufficient to specify the mesh in \colorbox{mygray}{\lstinline|/computation|}. 
The mesh should be ideally in a common format, e.g., msh or STL. However, STL can only define surface meshes and no volumetric ones, so for complicated structures, such as ones consisting of multiple materials, other formats can be used.  To simplify access to the mesh for the user, a \emph{softlink} \colorbox{mygray}{\lstinline|/mesh|} to the mesh location can be provided at the top level, if possible. In exceptional cases, when there is no possibility to describe the mesh using a common format without losing information about the mesh or deforming it, the specific list of mesh parameters datasets can be stored inside a \emph{group} \colorbox{mygray}{\lstinline|mesh|}. A unit of length has to be added to the mesh definition as an \emph{attribute} \colorbox{mygray}{\lstinline|unit|}. The \emph{attributes} \colorbox{mygray}{\lstinline|name|}, \colorbox{mygray}{\lstinline|description|}, and \colorbox{mygray}{\lstinline|keywords|} can provide additional information.



\subsection{Embedding}
The \emph{group} \colorbox{mygray}{\lstinline|/embedding|} is a stand-alone \emph{group} describing the embedding medium that has the same structure as the \emph{group} \colorbox{mygray}{\lstinline|/material|}. Anisotropic or non-reciprocal materials are not allowed for the embedding. Chiral materials are only allowed with the helicity basis for the T-matrix.

\subsection{Computation}

To reproduce the data of the T-matrix, this section should contain information on the way how it was computed. Because there are many ways of obtaining a T-matrix, this section should be adapted to different situations. The used software and its version are specified in the \emph{attribute} \colorbox{mygray}{\lstinline|software|}. It is required to add all the used software in a comma-separated string, including the version, and in particular, the repository~\cite{data_format}, where the template scripts are located, is to be defined in the form \enquote{\texttt{tmatrix\_data\_format=vx.x.x}}. Because differences between implementations of the HDF5 wrappers can occur, it is required to specify the program and its version used to create the HDF5 file in the same way as for other software, e.g., \enquote{\texttt{h5py=version}}. In case several scripts realize different approaches to compute the T-matrix using the same external software, or there is an external repository hosting the original scripts, the specific links can be added in the \colorbox{mygray}{\lstinline|reference|} \emph{attribute} of \colorbox{mygray}{\lstinline|/computation|}. The \emph{attribute} \colorbox{mygray}{\lstinline|method|} of \colorbox{mygray}{\lstinline|/computation|} describes the computational technique that the software implements. The best approach is to include in a comma-separated string both the abbreviation and the full name of the method. 

Additionally, \colorbox{mygray}{\lstinline|/computation|} should contain the files needed to reproduce the data in a dedicated \emph{group}  \colorbox{mygray}{\lstinline|/computation/files|}, e.g., full Python or other programming language script source codes.
To simplify the search for parameters used in the specific computation, a \emph{group}  \colorbox{mygray}{\lstinline|/computation/method_parameters|} includes as \emph{datasets} all the specific numerical values, with the names of the \emph{datasets} following the ones used by the software as closely as possible. It is important for consistency to use the parameter names from the template scripts of the repository. The parameters  typically include information about mesh discretization and accuracy.

The subtle aspect, which can become significant for the analysis of the data, is the question of which entries of the T-matrix do not include any numerical inaccuracies. Some computational methods can leverage symmetries of the objects, such that, for example, the response of a rotationally symmetric 3D object can be reduced to solving a 2D problem. Then, the entries that are zero due to symmetry automatically are set in the final T-matrix and are not the result of a simulation. To store this information, we suggest to add 
a mask in a \emph{dataset}  \colorbox{mygray}{\lstinline|/computation/analytical_zeros|} of the same shape as the T-matrix, where 0 stands for analytical zero entries, and 1 is set otherwise. 
This specification is different from the keywords of the T-matrix, where one can specify the symmetry of the object since the rotationally symmetric object can still be computed with a method that performs full 3D computation and produces only approximate zeros. Finally, it is possible to add the mesh used in the computation in this group, as described in the geometry section.

\part{Generation and validation}
This part describes methods to generate files according to the standard and validate them. 

\section{Generating files}
We stress at first that we provide files in a dedicated repository~\cite{data_format} that essentially consists of scripts that can be run to (a) compute T-matrices and (b) to assemble them in a way into files so that these files inherently agree with the specifications of the data format. So in a nutshell, if you use these files, you will obtain results that intrinsically agree with all the requirements. We describe a few of these methods in the following, but the developments are rapid and with time passing, we expect a larger number of tools to be adapted to compute and store these T-matrices in the required data format. Therefore, the list of current tools can only be considered as a snapshot.  

Incorporating the T-matrix formalism with other numerical methods and different software implementations is of further great use to the community. We encourage everybody to consider adapting their tools so that they can be used for the same purpose and making these files available to the public via the mentioned repository~\cite{data_format}. 

A first sanity check for the correct implementation is to demonstrate that the computed T-matrix indeed is constructed such that it stores the response of the object for all illumination directions. It is known that the T-matrix of an object itself does not depend on the incident illumination, while the scattering coefficients do. Therefore, an additional check is recommended for the scattering cross-section of the object at a specific illumination using the full-wave simulation software and the computed T-matrix. This also will increasingly apply to machine learning tools for synthetic data generation. Over time, example use cases for synthetic data generation tools can be added. These tools may systematically vary parameters to create diverse and comprehensive datasets, enhancing the training of machine learning models and closing gaps in the available data.

To add a new method to the repository, the files to produce the T-matrix should be submitted as a pull request to the repository together with benchmarking results for some basic examples. They are manually inspected afterward. Once this verification has been done, the new codes to generate T-matrices will be made available to the public via the repository. Ideally, a small description is provided in agreement with the documentation for the already established methods, as outlined in the following. 

In the following, we demonstrate the extraction of the T-matrix using various software.  
\subsection{JCMsuite}
\label{ch:jcm}

The program JCMsuite has the built-in capability to illuminate objects with vector spherical waves and calculate the decomposition of the scattered fields. Thus, it is well suited to compute T-matrix coefficients~\cite{santiago2019decomposition}. It uses the finite element method (FEM) to compute the scattering response and can be applied to arbitrary shapes. For illumination of a scattering object by multiple sources of the same frequency, it allows us to generate multiple independent solutions at the computational cost of a single solution by reusing the inverted system matrix~\cite{Burger2013al}. A similar approach for T-matrix extraction from commercial software (Finite Element based
HFSS) using different incident angles has been developed by Huang {\it et al.} \cite{huang2017hybrid}. It also provides the full definition of the bianisotropic tensor. However, the decomposition is done in a parity basis, so the embedding medium has to be achiral.

Generally, a simulation with JCMsuite is controlled by several files. These are, at minimum, a project, a source, a material, and a layout file. Combining these files with a script using MATLAB or Python to perform, e.g., automatic parameter sweeps is possible. Since all these files are text-based, including full documentation of the simulation setup in the HDF5 file is simple.

In our examples, the first three files, the project, sources, and material file, stay mostly the same. The project file typically looks like
\begin{lstlisting}[language=python,caption=project.jcmpt]
Project {
  Electromagnetics {
    TimeHarmonic {
      Scattering {
        FourierModeRange = [0, %(degree_max)e]
        Accuracy {
          Precision = %(precision)e
          Refinement {
            MaxNumberSteps = %(max_refinements)e
            Strategy = PAdaptive
          }
          FiniteElementDegree = %(fem_degree)e
        }
      }
    }
  }
}

PostProcess {
  MultipoleExpansion {
    FieldBagPath = "project_results/fieldbag.jcm"
    OutputFileName = "project_results/vsh.jcm"
    MultipoleDegree = %(degree_max)e
  }
}
\end{lstlisting}
and includes a general setup of the type of calculation to perform, some settings of numerical parameters which control the solution accuracy, and the post-process to perform the decomposition. In this example, several parameters are taken from the Python script. In the configuration language of JCMsuite, they are indicated by the percent symbol, a variable name in brackets, and a variable type indicator letter. Besides parameters that define the accuracy of the FEM calculation, the maximal multipole order has to be set in the variable \verb|degree_max|. 

Next, the source file contains the definition
\begin{lstlisting}[language=python,caption=sources.jcmt]
<?
for degree in range(1, keys['degree_max'] + 1):
    for order in range(-degree, degree + 1):
        for pol in 'MN':
            keys['degree'] = degree
            keys['order'] = order
            keys['pol'] = pol
            ?>

SourceBag {
  Source {
    ElectricFieldStrength {
      VectorSphericalWaveFunction {
        Coefficient = 1
        Lambda0 = %(lambda0)e
        MultipoleDegree = %(degree)i
        MultipoleOrder =  %(order)i
        Type = %(pol)s
      }
    }
  }
}

<?
?>
\end{lstlisting}
which uses simply a loop over vector spherical waves up to the maximum multipole order. The correct wavelength must also be set in the variable \verb|lambda0|. The materials file includes a loop over all material parameters set in the Python script.
\begin{lstlisting}[language=python,caption=materials.jcmt]
<?
for mat, er in enumerate(keys['epsilon']):
    keys['permittivity'] = er
    keys['mat_id'] = mat + 1
    keys['mat_name'] = f'Material_{mat}'
    ?>

Material {
  Name = "%(mat_name)s"
  DomainId = %(mat_id)e
  RelPermittivity = %(permittivity)e
}

<?
?>
\end{lstlisting}
At least the relative permittivity has to be set. Relative permeability and the chirality parameter are optional. The first material in the list is taken as the embedding material, which is associated with the domain ID 1 within the setup of JCMsuite used here.

Finally, the layout file contains the description of the geometry. In the example, we consider a single sphere. Due to the rotational symmetry, it is possible to restrict the FEM calculation to a two-dimensional domain. According to the domain IDs in the materials file, we set the background to domain ID 1 and then include the (semi-)circle for the object.

\begin{lstlisting}[language=python,caption=layout.sphere.jcmt]
Layout2D {
  CoordinateSystem = Cylindrical
  UnitOfLength = %(uol)e
  MeshOptions {
    MaximumSideLength = %(maxsl)e
  }
  Objects {
    Parallelogram {
      Priority = -1
      DomainId = 1
      Width = %(domain_radius)e
      Height = %(domain_z)e
      Port = West
      Boundary {
        Class = Transparent
      }
    }
    Circle {
      DomainId = 2
      Priority = 1
      Radius = %(radius)e
      MeshOptions {
        MaximumSideLength = %(object_maxsl)e
      }
    }
  }
}
\end{lstlisting}

The whole computation is controlled from a Python script. While the script itself is a somewhat longer, for the setup and subsequent run of the calculation, the relevant sections are the definitions of the keys used to fill the open variables in the JCMsuite files


\begin{lstlisting}[language=python,caption=Snippet from Python script sphere\_jcmsuite.py -- defining the keys,firstline=21,lastline=37,firstnumber=18]
    keys = {
        "uol": 1e-9,  # unit of length (1 = m, 1e-9 = nm)
        "epsilon": [1, 4],  # First entry embedding
        "mu": [1, 1],
        "radius": 100,
        "degree_max": 5,
    }
    keys_method = {
        "domain_radius": 125,
        "domain_z": 250,
        "maximum_sidelength_domain": 10,
        "maximum_sidelength_object": 5,
        "precision": 1e-7,
        "max_refinements": 3,
        "fem_degree": 2,
    }
    keys.update(keys_method)
\end{lstlisting}
and the loop to start all computations.

\begin{lstlisting}[language=python,caption=Snippet from Python script sphere\_jcmsuite.py -- loop,firstline=50,lastline=63,firstnumber=50]
    jobids = []
    for i, freq in enumerate(freqs):
        keys["lambda0"] = c0 * keys["uol"] / freqs[i]
        jobid = jcmwave.solve(
            "project.jcmp",
            keys=keys,
            working_dir=f"{working_dir}/job{i}",
            jcmt_pattern=jcmt_pattern,
        )
        jobids.append(jobid)
    results, logs = jcmwave.daemon.wait(jobids)
    jcmwave.daemon.shutdown()

    tmats, _, ls, ms, pols = jcm_tools.extract_tmatrix_jcm(results)
\end{lstlisting}

In the final line of the listing, we extract the T-matrices from the results using a custom function. In the last lines of the original script, the data is stored using the module \verb|tmatrix_tools.py| that provides a collection of functions to create a standard conforming T-matrix file.

\subsection{COMSOL}
The same methodology can be implemented in COMSOL Multiphysics, one of the most used FEM software programs. 
To demonstrate the capabilities of COMSOL, we will concentrate on the case where symmetry considerations can cut down the calculation time. The simplest example would be a sphere, which has rotation symmetry. The full Java script version was made available to the public repository~\cite{data_format} as well. The compiled scripts, with commented-out
last lines to avoid immediate running, can then be opened and modified manually in the GUI. For demonstration purposes, the screenshots of COMSOL's GUI are included in this section.

\begin{figure}[h!!]\label{geom_comso}
  \centering
  \includegraphics[width=1\linewidth]{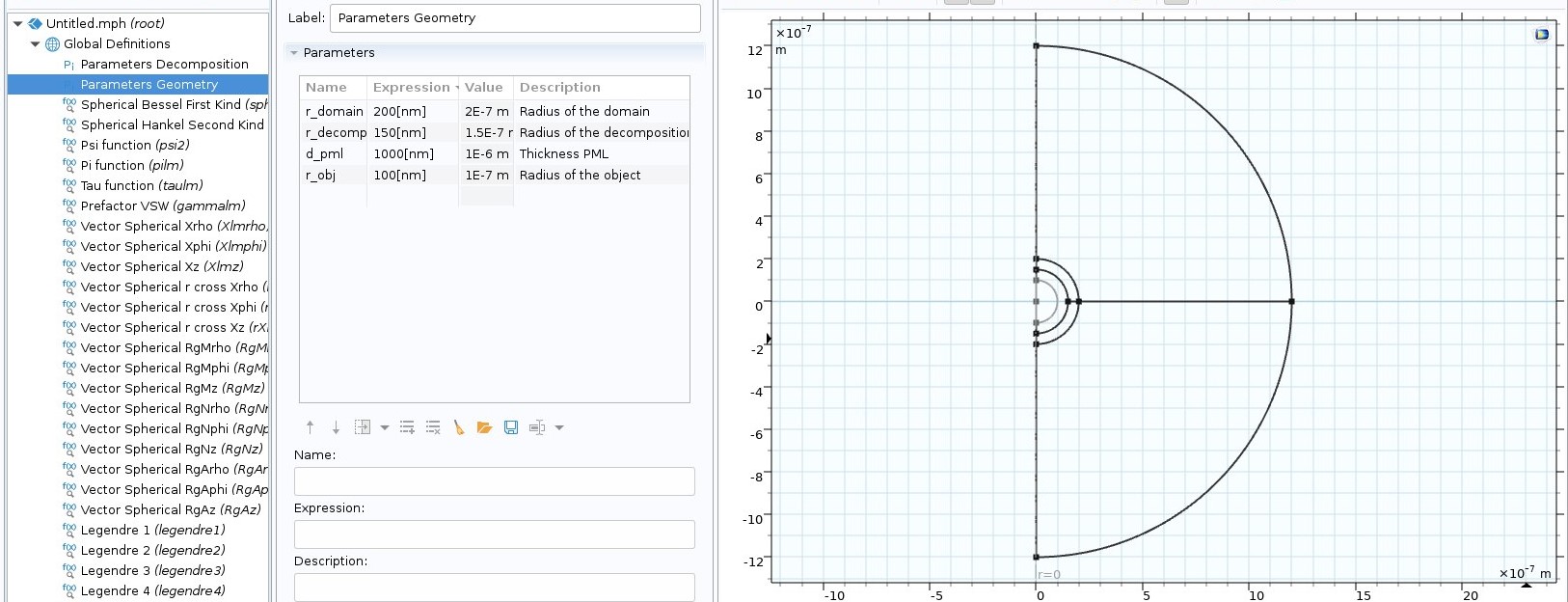}
  \caption{COMSOL: Specification of the geometry of the problem in the GUI}
  \label{geom_comsol}
\end{figure}

The first step is to define the parameters, and here, the additional circumscribing sphere is introduced, defined as the decomposition radius. This should fully include the object and it is the surface where the decomposition of the fields into VSWFs will be performed. The domain has to be larger than the decomposition radius and is followed by the perfectly matched layer (PML)~\ref{geom_comsol}. Rotationally symmetric examples require additional care in the choice of the domain size, such that it is comparable to the wavelength~\cite{gladyshev2024}. Under parameter decomposition, the maximum desired multipole order for the incident waves and scattered waves is defined. The customized part of the model is the function definitions for the incident waves. To obtain a T-matrix of an arbitrary object, it should be illuminated with VSWFs up to the desired multipole order. Therefore, definitions of Legendre polynomials, VSWFs, Bessel, and Hankel functions are necessary. 
As an additional note, the associated Legendre polynomials are implemented in versions starting from COMSOL 5.5., for the earlier versions one can use  manually implemented functions up to degree 5, such that (3D) T-matrix computations are possible only up to degree 4. 

\begin{figure}[h!!]
  \centering
  \includegraphics[width=1\linewidth]{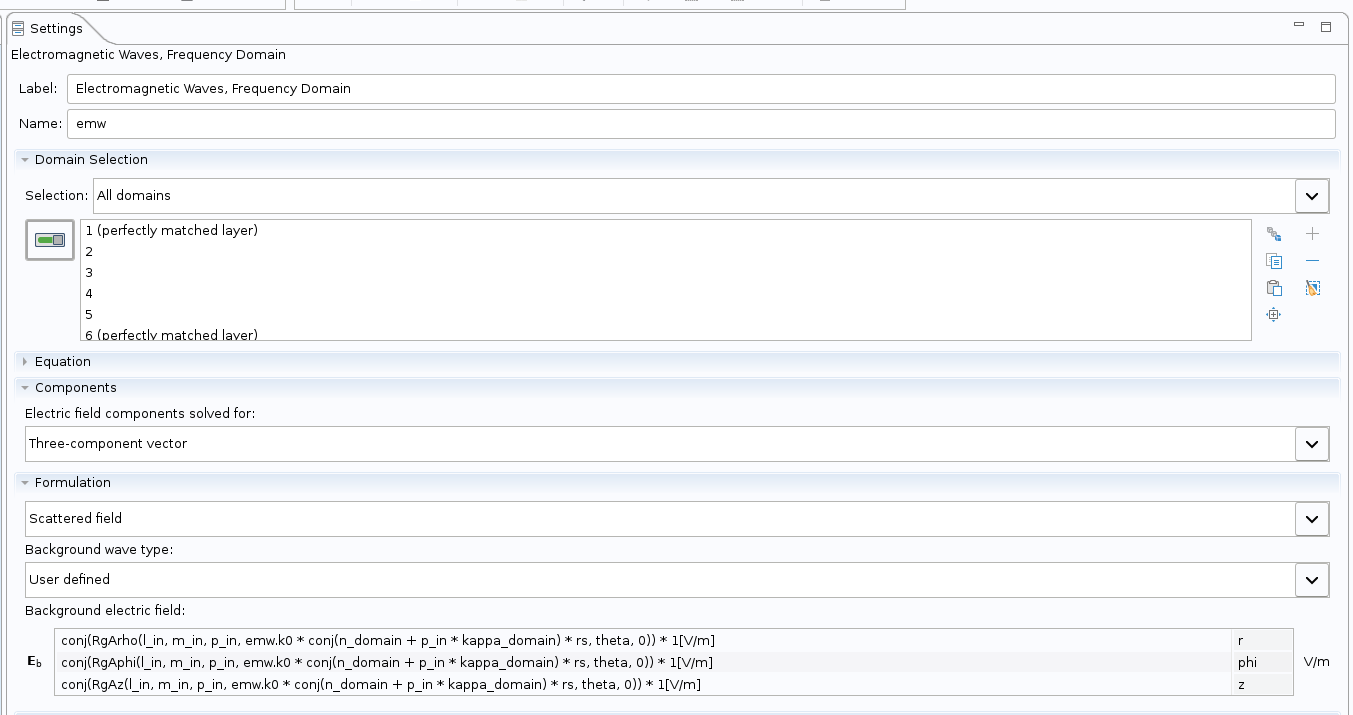}
  \caption{COMSOL: Background electric field definition in the GUI}
  \label{ns}
\end{figure}

In the definition of the geometry of the object, half of the circle is given, and the region of the circumscribing sphere is technically included in the domain, such that the domain has a few layers. The chirality parameter can be set to a non-zero value. The Variables Decomposition includes the integration  formulas of VSWFs and scattered fields  on the circumscribing sphere. In this model, the scattered field formulation is used, so the resulting field is a superposition of the background field and the scattered field. The definition of the background electric field, the solution of Maxwell's equations in the absence of the scatterer, is making use of the
special functions defined previously (see Fig.~\ref{ns}). An alternative total field formulation is also possible. Next, we choose axial symmetry for all boundaries. The equation view reveals the whole set of expressions. The mesh size can be controlled as well.

To finally run the computation, two studies are prepared. The first study computes the solution to the scattering problem. One can select a range of frequencies for the calculation, in addition to the nested parameter sweeps for the order and polarization of the vector spherical waves. The second study post-processes these results, evaluates the necessary integrals on the decomposition surface, and computes the T-matrix entries. The results obtained from the second study are in the form of a list of variables \enquote{$\mathrm{ap}$}, and \enquote{$\mathrm{am}$}, which stand for positive and negative helicity. In the final format used for the storage, one has to interleave these two matrices of coefficients. The Python script to convert the results to \texttt{.tmat.h5} format is provided as well in~\cite{data_format}.

\subsection{ONELAB}
In this section, we demonstrate the retrieval of T-matrices with the open-source FEM solver ONELAB~\cite{geuzaine2013onelab}. It is based on the mesh generator Gmsh~\cite{dular1997getdp} and the finite element solver GetDP~\cite{geuzaine2009gmsh}. The underlying theoretical treatment and the numerical implementation details are reported in~\cite{demesy2018scattering}. In line with the previous sections, the VSWFs are implemented directly as the illumination fields. The simulations are conducted using a scattered field formulation of FEM, which makes it possible to locate the illumination sources within the scatterer. In this implementation, the final integration with VSWFs is performed on a spherical surface. As exemplified in the snippet~\ref{onelabsrc} from a single sphere T-matrix computation code, after the initialization of the ONELAB client and definition of the parameters, the Gmsh and GetDP programs are run as subclients over a range of wavelengths. The results are stored in a particular ordering using the \texttt{\_get\_postprocess} function. Then, the \texttt{tmatrix\_tools.py} utility module is used to store the data in HDF5 file.    
\lstinputlisting[label={onelabsrc}, language=python, caption=Snippet from sphere\_onelab.py,firstline=110,lastline=155]{examples/onelab/sphere\_onelab.py}
The source files are provided in~\cite{data_format}. 

\subsection{nanobem}
\textit{nanobem} is an open-source MATLAB toolbox, aimed among other functionalities at solving Maxwell's equations using the boundary element method (BEM). In this implementation, the method is based on a Galerkin scheme with Raviart-Thomas basis functions~\cite{hohenester2020nano}. The material of the scatterers is assumed to be linear, homogeneous, and local, separated between different media via abrupt interfaces. For a detailed review of the toolbox, see \cite{hohenester2022nanophotonic, hohenester2024nanophotonic}. In the following, we solely demonstrate how the construction of a T-matrix is realized using the toolbox. 

Unlike in the FEM method, the computation of the field in the BEM method is performed at the surface of the object. For this reason, the problem is reformulated, and the previously defined coefficients include integration not on the circumscribing sphere, but on the surface of the object. The surface boundary is denoted as ${\partial \Omega}$, and the normal to the surface boundary is $\mathbf{\hat{n}}$. The representation formula allows us to compute the field at position $\mathbf{r}$ by surface integration:
\begin{equation}
\mathbf{E}_\mathrm{sca}(\mathbf{r}) = \oint_{\partial \Omega} \{\mathrm{i} \mu \omega G(\mathbf{r, r'}) \cdot \mathbf{\hat{n}} \times \mathbf{H}(\mathbf{r'}) - [ \nabla \times G(\mathbf{r, r'})] \cdot \mathbf{\hat{n}} \times \mathbf{E}(\mathbf{r'})\} {\mathrm d}S'\, ,
\end{equation}
where $G(\mathbf{r}, \mathbf{r'})$ is the dyadic Green's function expansion in the basis of transverse vector spherical waves (Eq.~7.3.40 of~\cite{chew1999waves}):   
\begin{equation}
G(\mathbf{r}, \mathbf{r'}) = \mathrm{i}  k \sum_{l, m}{\left( \mathbf{M}_{lm}(\mathbf{r})  \mathbf{M}^\dagger_{lm}(\mathbf{r'}) + \mathbf{N}_{lm}(\mathbf{r})  \mathbf{N}^\dagger_{lm}(\mathbf{r'})  \right)}\, .
\end{equation}
For the T-matrix calculation we always have $r>r'$, therefore $\mathbf{M}_{lm}(\mathbf{r})$, $\mathbf{N}_{lm}(\mathbf{r})$ contain the spherical Hankel, while $\mathbf{M}^\dagger_{lm}(\mathbf{r'})$, $\mathbf{N}^\dagger_{lm}(\mathbf{r'})$ the spherical Bessel functions.
From this equation, the T-matrix elements can be computed by (i) illuminating the scatterer with vector spherical waves, and (ii) by integrating the inner product between the induced tangential electromagnetic fields and $\mathbf{M}^\dagger_{lm}(\mathbf{r'})$, $\mathbf{N}^\dagger_{lm}(\mathbf{r'})$ over the boundary of the scatterer.

 
With the numerically implemented formulas, the computation of the T-matrix can be exemplified in the following code snippet.            
\begin{lstlisting}[language=MATLAB, caption=Snippet from  demomulti01.m: T-matrix computation for a sphere using BEM ,linerange={4-32},firstnumber=1]
mat1 = Material( 1, 1 );
mat2 = Material( 9, 1 );
%  material vector
mat = [ mat1, mat2 ];

%  discretized sphere boundary with 256 vertices
diameter = 160;
p = trisphere( 256, diameter );
%  boundary elements with linear shape functions
tau = BoundaryEdge( mat, p, [ 2, 1 ] );     

%  wavenumber of light in vacuum
k0 = 2 * pi / 1000;
%  BEM solver and T-matrix solver
lmax = 4;
bem = galerkin.bemsolver( tau, 'order', [] );
tsolver = multipole.tsolver( mat, 1, lmax );
%  T-matrix
sol = bem \ tsolver( tau, k0 );
t1 = eval( tsolver, sol );

%  additional information for H5 file
info = multipole.h5info( tau );
info.name = 'Sphere';
info.description = 'Single sphere and single wavelength';
info.matname = [ 'Embedding medium', 
'Dielectric medium of sphere' ];
%  save T-matrix
fout = 'tmatrix_sphere.tmat.h5';
\end{lstlisting}
In the first few lines, the parameters of the object are defined, including the material and geometry. The first material in the list corresponds to the background medium, and the second to the scatterer. The mesh triangulation is specified as the next step, here a triangular boundary element shape is selected. Next, for the desired wavelength and number of multipole orders indicated, the method \texttt{galerkin.bemsolver} is used to initialize the solver object, and \texttt{multipole.tsolver} initializes the T-matrix object. The variable \texttt{sol} contains the solution obtained by the BEM method, which includes the tangential electric and magnetic fields. The \texttt{eval} function uses the computed fields to calculate the T-matrix entries. The T-matrix is finally stored in the HDF5 file together with the required metadata. Further information can be obtained from the help pages of the toolbox.

\subsection{ADDA}
ADDA~\cite{ADDAR1} is a robust open-source implementation of the DDA~\cite{ADDAR2,ADDAR3}. It can handle particles of any shape and composition, including non-spherical and inhomogeneous particles, as well as particles with material anisotropy. It is optimized for high performance, supporting distributed memory clusters, multi-core processors, and GPU acceleration. This enables the analysis of large particles with high refractive indices compared to other DDA-based software tools. This acceleration is particularly important, as calculating the T-matrix requires simulations for multiple illuminations.

Although illumination with VSWFs is currently not available in ADDA, it is possible to retrieve the T-matrix by determining the scattered field from multiple plane wave illuminations. To that end, we adopt the approach proposed by Fruhnert {\it et al.} \cite{ADDAR5}, which involves the decomposition of each incident field and the corresponding scattered fields into VSWFs.
The T-matrix is then obtained by solving an inverse problem that relates the coefficients of the scattered and incident field expansions into adequate VSWFs for $K$ different illuminations:
\begin{equation}
\label{Eq:ADDA1}
\left(\mathbf{p}^{(1)} \mathbf{p}^{(2)} \cdots \mathbf{p}^{(K)}\right)=\mathbf{T}_K \cdot \left(\mathbf{a}^{(1)} \mathbf{a}^{(2)} \cdots \mathbf{a}^{(K)}\right)\, .
\end{equation}

The initial method proposed in Ref.~\cite{ADDAR5} expands the fields on a spherical shell enclosing the analyzed nanoparticle. In the following, we rely on the scattered far-field instead, as this quantity is more readily available in ADDA and is generally easier to compute.
The core quantity in ADDA is the polarization of each dipole ($\mathbf{P}_i$) that is calculated self-consistently based on the dipole polarizability ($\overline{\boldsymbol{\alpha}}_i$) and Green's tensor ($\overline{\mathbf{G}}_{i j}$) for a given incident field on each dipole ($\mathbf{E}_i^{\text{inc}}$)~\cite{ADDAR2}:
\begin{equation}
\label{Eq:ADDA2}\overline{\boldsymbol{\alpha}}_i^{-1} \mathbf{P}_i-\sum_{j \neq i}\overline{\mathbf{G}}_{i j} \mathbf{P}_j=\mathbf{E}_i^{\text{inc}}\, .
\end{equation}
Once the polarization is determined, the scattered far-field is obtained as:
\begin{equation}
\label{Eq:ADDA3}
\mathbf{E}_{\text{sca}}(\mathbf{r})=  \frac{\exp \left(\mathrm{i} k r\right)}{-\mathrm{i} k r}\mathbf{F}(\hat{\mathbf{r}})\, ,
\end{equation}
where the scattering amplitude $\mathbf{F}$ depends only on the scattering direction $\hat{\mathbf{r}}=\mathbf{r}/r$
\begin{equation}
\label{Eq:ADDA4}
\mathbf{F}(\hat{\mathbf{r}})= -\mathrm{i}k^3 (\overline{\mathbf{I}}-\hat{\mathbf{r}} \otimes \hat{\mathbf{r}}) \sum_i \mathbf{P}_i \exp \left(-\mathrm{i} k \mathbf{r}_i \cdot \hat{\mathbf{r}}\right)\, .
\end{equation}
Here, $\overline{\mathbf{I}}$ is the identity tensor and $\hat{\mathbf{r}} \otimes \hat{\mathbf{r}}$ projects any vector on $\hat{\mathbf{r}}$. Note that Eqs.~\ref{Eq:ADDA2} --\ref{Eq:ADDA4} are based on the Gaussian-CGS system of units, as the one employed in ADDA. However, that is not relevant for the finally computed quantities, like the T-matrix and cross-sections.

To find the expansion coefficients, we leverage the far-field limit of the VSWFs (see Eq.~\ref{Eq:C3}) and cancel the common dependence on $r$ in these functions and $\mathbf{E}_{\text{sca}}(\mathbf{r})$. Then, Eq.~\ref{Eq:ADDA4} transforms into:
\begin{equation}
\mathbf{F}(\hat{\mathbf{r}}) = -k^2\sum_{l=1}^{\infty} \sum_{m=-l}^{l} (-\mathrm{i})^{l}\left[ p_{lm}^{\text{e}} \mathbf{Y}_{lm}(\hat{\mathbf{r}}) + p_{lm}^{\text{m}} \mathbf{X}_{lm}(\hat{\mathbf{r}}) \right]\, ,
\end{equation}
where the normalized vector spherical harmonic $\mathbf{X}_{lm}(\hat{\mathbf{r}})$ is defined in~\ref{app:normalization} and its counterpart $\mathbf{Y}_{lm}(\hat{\mathbf{r}})$ is defined as $\mathbf{Y}_{lm}(\hat{\mathbf{r}})=\mathrm{i} \hat{\mathbf{r}} \times \mathbf{X}_{lm}(\hat{\mathbf{r}})$. This decomposition together with mutual orthogonality and normalization of the harmonics leads to the following calculation of the expansion coefficients:
\begin{equation}\label{Eq:ADDA5}
\begin{aligned}
p_{lm}^{\text{e}}&=-\frac{\mathrm{i}^l}{k^2} \int_0^{2 \pi} \int_0^\pi \mathbf{F}(\theta,\phi)\mathbf{Y}^*_{lm}(\theta,\phi) \sin{\theta} \, \mathrm{d} \theta \, \mathrm{d} \phi\, , \\
p_{lm}^{\text{m}}&=-\frac{\mathrm{i}^l}{k^2}\int_0^{2 \pi} \int_0^\pi \mathbf{F}(\theta,\phi)\mathbf{X}^*_{lm}(\theta,\phi) \sin{\theta} \, \mathrm{d} \theta \, \mathrm{d} \phi\, . \\
\end{aligned}
\end{equation}

To form an inverse problem of Eq.~\ref{Eq:ADDA1}, one also needs the incident field coefficients. For the plane-wave illumination, those can be obtained analytically. To that end, we use the expressions from \cite{Doicu2006} with changes according to different definitions of VSWFs.

The code implementing the T-matrix calculation with ADDA is written in Python and hosted on GitLab~\cite{addatmatrix}. 
The installation process is simplified by automatically setting up ADDA and designed to be cross-platform, having been tested on Windows and Linux environments, specifically under Python 3.11.

At its core, this implementation uses simple trapezoid integration for the numerical tasks, which, while straightforward, is effectively enhanced with Numba to significantly boost computational speed. To solve the T-matrix, the code calculates the pseudoinverse of the matrix that represents the incident field coefficients.

Calculation of the T-matrix with the provided code is executed either using a simple Command-Line Interface (CLI) that mimics the standard ADDA CLI or by writing a custom Python script using the dedicated Python package. With the CLI, the approach is designed to be accessible for users familiar with ADDA's conventional interface. The main argument is, then,  a string with a set of standard ADDA parameters. For example, here is a command executing the calculation of the T-matrix of a sphere with refractive index of 2, diameter 200 nm, at a wavelength of 500 nm:
\begin{lstlisting}[language=sh,caption=T-matrix generation using ADDA CLI ]
./addatmatrix_cli.py tmatrix --adda_string "-shape sphere -m 2 0  -size 200 -lambda 500"
\end{lstlisting}

It is a good practice to also specify the number of illuminations $K$, size of the T-matrix, and number of scattering angles for quadratures in Eq.~\ref{Eq:ADDA5}. However, some default values are present in the script.

Using the provided Python package simplifies scripting and integration into Python-based projects. For example, here is the same simulation as above:
\begin{lstlisting}[language=python,caption=T-matrix generation using Python code]
from addatmatrix.data import write_t_matrix_data
from addatmatrix.t_matrix import t_matrix

output_path = "sphere.tmat.h5"
adda_string = "-shape sphere -m 2 0"
tmatrix_data, metadata, indices = t_matrix(adda_string,wavelength=500)
write_t_matrix_data(output_path, tmatrix_data, indices, metadata)
\end{lstlisting}

Looking towards future improvements, the code could benefit from FFT acceleration of far-field scattering calculation~\cite{ADDAR4} and more advanced integration methods. Alternatively, the far-field scattering can be skipped altogether by direct computation of VSWF expansion coefficients from the dipole polarizations. The latter resembles the translation-addition of VSWFs, for which fast algorithms exist as well. Moreover, the incident illumination can also be changed to VSWFs - this will eliminate the calculation of pseudo-inverse, making the whole approach robust for particles larger than the wavelength. These enhancements would further solidify the code's utility and performance, making it an even more powerful tool for researchers and engineers working in nanophotonics and other fields.

\subsection{MEEP}
MEEP stands for ``MIT Electromagnetic Equation Propagation'' and solves Maxwell's equations via the finite difference time domain method (FDTD)~\cite{oskooi2010meep}.    
Since the T-matrix connects fields in the frequency domain, the Fourier transform is part of the post-processing. A single simulation in time domain allows obtaining T-matrices at multiple frequencies. Similarly to the previous example, the complication arises from the fact that direct illumination with a VSWF is not an option in MEEP. Instead, plane wave illuminations are used, and a decomposition into spherical waves is performed afterward. Specifically, the plane waves in the following form are used:
\begin{equation}
    \mathbf{E}_\mathrm{in}(\mathbf{r}, t) = \mathbf{E}_{0}  e^{i(\mathbf{k_{0}\cdot r} - \omega t)} g(\mathbf{k_{0}\cdot r} - \omega t)\, ,
\end{equation}          
with a suitable envelope function $g(\mathbf{k_{0}\cdot r} - \omega t)$. The spectrum that is launched into the system is dictated by the temporal width of the envelope function. The total simulations have to cover all the angles of incidence and all the polarizations as fully as possible. The approach is to generate an equally distributed grid of points on a sphere following the ``Fibonacci Sphere'' algorithm and in this way define the incident angles, similarly to the approach described previously with ADDA. For each incident direction, fields with two orthogonal polarizations have to be defined. The intricate detail is the choice of sufficient illumination sources to retrieve the correct scattering response for any illumination just by multiplying the incident field with a computed T-matrix. A T-matrix of size $N\times N$ requires a minimum number of $N_\mathrm{min}=2(l_\mathrm{max}+2)l_\mathrm{max}$ equations. However, for a more stable solution, it is recommended to perform more simulations, typically $N_{\text{sim}} = 2 N_\mathrm{min}$. In the current implementation, the simulation is not performed with a rotated plane wave, but with a rotated object. Rotation of plane waves is not desired, since the oblique plane waves are not fully absorbed in perfectly matching layers, and the wavefront is thus slightly distorted. Before assembling the T-matrix, the fields are rotated back. After the transformation to the frequency domain is done, the scattering coefficients are calculated using the formula from \cite{santiago2019decomposition}:
\begin{equation}\label{Eq:santiago}
  \{a, b\}_{lm} = -ik \int_{\Gamma} \mathbf{dS} \left[ (\nabla \times \mathbf{E}_\mathrm{sc}) \times \{ \mathbf{N}^{(1)*}_{lm}, \mathbf{M}^{(1)*}_{lm} \} -k  \{ \mathbf{M}^{(1)*}_{lm}, \mathbf{N}^{(1)*}_{lm} \} \times \mathbf{E}_\mathrm{sc} \right]\, ,
\end{equation}
here ${\Gamma}$ is an arbitrary surface enclosing the scatterer. The surface of a  cuboid in this case is beneficial, since the grid is rectangular.  

In the following, the sequence of calls to different functions is shown in a few selected code snippets.  

\begin{lstlisting}[language=python,caption=main.py,linerange={69-71, 74},lastline=74]
def sphere() -> None:
    c = get_test_config()
    t = calculate_T(c)
    h5save(c, t, "test_sphere", "Test single sphere file")
    t_treams = treams.TMatrix.sphere(
    c.l_max, c.f_cen * 2 * np.pi, c.params["radius"], [treams.Material(c.material), treams.Material(c.embedding)]
    )
    t_treams = t_treams.expand(treams.SphericalWaveBasis.default(c.l_max))
    t_treams = t_treams.changepoltype()

    t = t[:, : t_treams.shape[-2], : t_treams.shape[-1]]
    plot_me_vs_treams(t, t_treams, c.path_output)
    delta = delta_treams_vs_me(t, t_treams)

if __name__ == "__main__":
    tetrahedron()
\end{lstlisting}

The parameters are uploaded from a preset configuration.
In the next code snippet, it can be seen that a few of them are set manually, while others are taking the default values.
\begin{lstlisting}[language=python,caption=Snippet from config\_factory.py,linerange={60-61, 64-74},lastline=268]
def get_test_config() -> Config:
    c = Config(
        resolution=30,
        sim_amount_mult=2,
        l_max=2,
        material=1.15,
        params={"radius" : 0.4},
        shape="sphere",
        eps_embedding = 1.,
        cpu_cores_per_simulation=SIZE,
    )
    return c
\end{lstlisting}
Resolution is one of the typical FDTD parameters, and the minimum accepted resolution in the code is one tenth of the wavelength. However, to reach converged results, a very high resolution is often required.  The parameter requiring some explanation is \verb|sim_amount_mult|. This factor is multiplied by the minimum number of simulations to define the total number of simulations to perform. As discussed previously, 2 is the optimal value. The maximum multipole order is set considering the size of the object and the wavelength. Further, parallelization parameters depending on the available resources is defined.

With the parameters fully specified, the computation flow proceeds to the function which performs the actual calculations. The \verb|calculate_T| function is called to start the (optionally) parallel execution of the simulations. Inside the inner function \verb|core_calc_T|, the MEEP simulations are set up and run. The 6 monitors are placed at the locations where the scattering coefficients will be computed using Eq.~\ref{Eq:santiago}, enclosing the scatterer in a cube. The distance to the monitors must be sufficiently large, such that the transient high-frequency fields excited when the source turns off decay sufficiently at such distance. The runtime can be directly specified, however, by default the simulation continues until the fields at the monitors are not changing with some tolerance after the sources were turned off. The incident fields are derived from a simulation of field propagation without the scatterer. After this, for all the incident angles and polarizations, the actual permittivity grid of the scatterer is rotated by the corresponding angle and the simulation is performed. The resulting fields are rotated back and Fourier transformed. The incident and scattered fields are expanded in VSWFs. Eventually, to obtain the T-matrix, the inverse problem is solved using the least squares method.

\begin{lstlisting}[language=python,caption=Snippet from the function \lstinline{core_calc_T} in scat.py, firstline=69,lastline=93]
    e_amp_inc = get_e_amp_from_e_in(c=c, source_data=source_data, sim_res=sim_res_inc)
    incident_matrix = []
    scatter_matrix = []
    rot_matrix = []
    for ii in sim_id_range:
        print(f"Simulation id = {ii}")
        rotation_data = get_persistent_sphere_angles(c=c, id=ii)
        rotated_epsgrid = rotate_eps_grid(eps_grid, rotation_data, c.eps_embedding)
        rot_array = np.array([rotation_data.theta, rotation_data.phi, rotation_data.alpha])
        rot_matrix.append(rot_array)        
        sim_res_sca = do_scattered_sim(
            id=ii,
            c=c,
            source_data=source_data,
            eps_grid=rotated_epsgrid,
            sim_res_inc=sim_res_inc,
        )
        incident_coefs = get_inc_coefs(
            c=c, source_data=source_data, rotation_data=rotation_data, e_amp=e_amp_inc
        )
        scatter_coefs = get_sca_coefs(
            c=c, rotation_data=rotation_data, sim_res=sim_res_sca
        )
        incident_matrix.append(incident_coefs)
        scatter_matrix.append(scatter_coefs)
\end{lstlisting}

As a final step, the function \verb|h5save| stores the T-matrix and the required metadata in a HDF5 file. In MEEP units, the speed of light is set to 1, which is considered when filling the metadata.

\subsection{SMARTIES}\label{smarties}

The original T-matrix method, devised by Waterman~\cite{waterman1965matrix}, introduced alongside a specific calculation scheme -- the Extended Boundary Condition Method (EBCM). This technique has strong analytical roots, requiring no meshing of the particle's volume or surface and, instead, computes T-matrix elements via analytical formulas which reduce to Mie theory for spherical particles~\cite{MishchenkoTL02}. For axial-symmetric particles, the method is remarkably efficient, as the matrix elements are obtained via simple one-dimensional integrals along the generatrix. The EBCM method is particularly popular for simple geometrical shapes, where it typically provides the fastest and most accurate way to calculate a T-matrix~\cite{MishchenkoTL02}.

SMARTIES~\cite{somerville2016smarties} is a MATLAB implementation of the EBCM to simulate the optical properties of oblate and prolate spheroidal particles, with comparable speed and accuracy as Mie theory for spheres. SMARTIES uses an improved algorithm that overcomes some loss of precision faced by EBCM in the case of large and elongated particles, and can routinely achieve numerical accuracy better than 10 digits. Although restricted to spheroids, SMARTIES may be useful to researchers seeking a fast, accurate and reliable tool to simulate the near-field and far-field optical properties of nonspherical particles, and can also appeal to other developers of light-scattering software seeking a very accurate benchmark for nonspherical particles with a challenging aspect ratio and/or refractive index contrast.

We provide below an example script (Listing~\ref{lst-smarties}) to output the T-matrix of a gold spheroid, used in the calculation of Fig.~\ref{fig-spheroid-dimer}. Further examples are available in the project's repository~\cite{Somerville_SMARTIES_2024}.

\begin{lstlisting}[label={lst-smarties},language=matlab,caption=Example usage of SMARTIES]
wavelength = (400:800)';
epsilon = epsAu(wavelength);
medium=1.33; % water

% spheroid semi-axes
% stParams.a=20; stParams.c=40;

 % simulation parameters
 stParams.N=3; stParams.nNbTheta=120;

 % additional options if required
 stOptions = {};

 % allocate 3D array for all results
 qmax = 2*(stParams.N*(stParams.N + 1) + stParams.N );
 tmatrix = zeros(length(wavelength), qmax, qmax);
 % loop over wavelengths
 for i=1:length(wavelength)
     stParams.k1=medium*2*pi/wavelength(i);
     stParams.s=sqrt(epsilon(i)) / medium;
     [stCoa, stT] = slvForT(stParams,stOptions);
     [T, u, up] = expand_tmat(stT, qmax); % tmatrix elements and indices
     ind = sub2ind(size(tmatrix),q*0+i, u, up);  linear index in 3D array
     tmatrix(ind) =  T(:,7) + 1i*T(:,8);
 end
 analytical_zeros = true(qmax,qmax);
 analytical_zeros(sub2ind(size(analytical_zeros), u, up)) = false;

 % data to export
 epsilon = struct(...
     'embedding', medium^2, 'particle', epsilon, 'embedding_name', 'H2O, Water', ...
     'embedding_keywords', 'non-dispersive', 'embedding_reference', 'constant', ...
     'material_name', 'Au, Gold', 'material_reference', 'Au from Johnson and Christy 1972',...
     'material_keywords', 'dispersive, plasmonic');

 geometry = struct('description',  'prolate spheroid', ...
     'shape', 'spheroid','radiusxy', stParams.a, 'radiusz', stParams.c);

 computation = struct('description', 'Computation using SMARTIES', ...
     'accuracy', 1e-10, 'Lmax', Lmax, 'Ntheta', Ntheta, ...
     'analytical_zeros', analytical_zeros);

 comments = struct('name', 'Au prolate spheroid in water',...
     'description', 'Computation using SMARTIES',...
     'keywords', 'gold, spheroid, ebcm', 'script', [mfilename '.m']);

 tmatrix_hdf5('smarties_example.tmat.h5', tmatrix, wavelength, epsilon,...
               geometry, computation, comments)
\end{lstlisting}

\begin{figure}

\centering{

\includegraphics[width=0.9\linewidth]{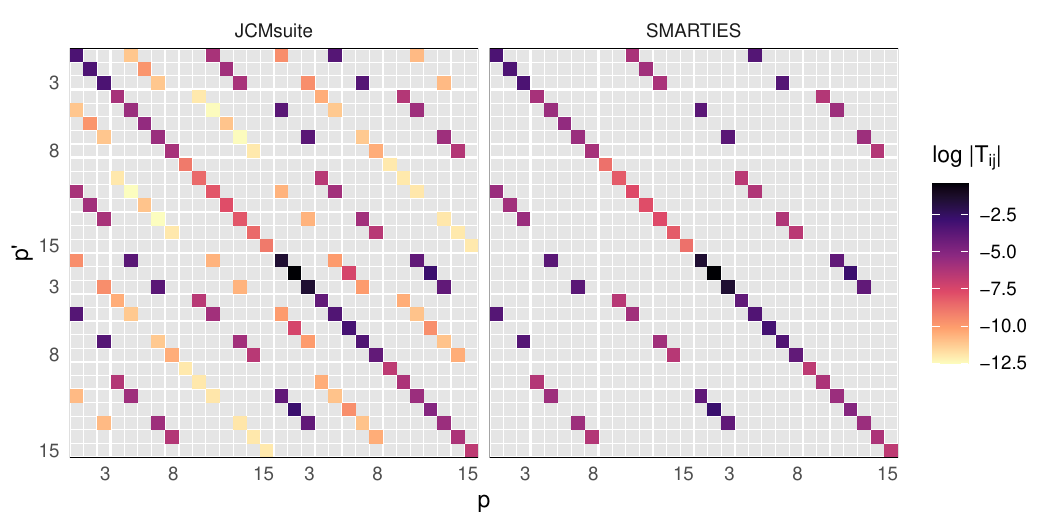}

}

\caption{\label{fig-spheroid-tmatrices}Comparison of T-matrices for a prolate Au spheroid immersed in water, calculated with FEM (JCMsuite, left), and the EBCM (SMARTIES, right). Both computations use a wavelength of 630~nm, and the spheroid semi-axes are 20~nm and 40~nnm. The color scale presents the modulus of the T-matrix elements (in log scale for clarity). The same pattern is observed for the dominant terms, but the FEM introduces a few non-zero elements due to the mesh description of the particle shape. In this plot, the native SMARTIES convention for T-matrix as 2x2 blocks is used.}

\end{figure}

Internally, SMARTIES stores T-matrix elements in a custom structure, as the particle symmetries (axial-symmetric and plane of symmetry) result in many T-matrix elements being exactly zero, and therefore not calculated~\cite{somerville2016smarties}. To convert to the \texttt{.tmat.h5} format, the function \texttt{expand\_tmat()} expands the full matrix by filling the missing elements with zeros, and also reorders indices to match the conventions described herein, from their original ordering as a \(2\times 2\) block matrix (Figure~\ref{fig-spheroid-tmatrices}).

\newpage
\section{Validation}
Before the final submission of the T-matrix file to the database, its compliance with the standard format developed and described in this document has to be proven. This includes technical aspects of the data representation and detecting physical inaccuracies pointing to faults in the computation method or discrepancies with the actual geometry setting. We stress that if the example files from the repository are used to generate the T-matrices, the requirements should be fulfilled, since they were inspected on individual basis. For the future database, the validation of the uploaded data will be performed using an automated script. 
\subsection{Formal validation}
Several points to comply with the formatting guidelines are important:
\begin{itemize}
    \item Nomenclature. The presence of all compulsory \emph{groups}, \emph{datasets}, and \emph{attributes} is systematically checked. Commonly used versions of the naming for the inputs produce suggestions to the user for a suitable alternative in the correct naming convention.
    \item Shapes and types. There is a correspondence of the stored parameters between each other, the size of the array of the T-matrix, and the size of the array of the modes defining polarization and multipole orders being one example. The types of the entries are also inspected. 

\end{itemize}

\subsection{Normalization conventions}
Different conventions regarding the vector spherical wave normalization are used in the literature. For the interchangeability of results calculated with different methods, it is important to adhere to one specific normalization, which we define in \ref{app:normalization}. As a check, a reference structure can be calculated with a custom method, and its T-matrix compared with the provided reference output.

\subsection{Physics constraints}\label{symmetries}
T-matrices can manifest physical constraints imposed by the properties of the investigated object such as symmetries~\cite{mishchenko1996t}. Combined with symmetries in lattices at which the scatterers are arranged, many physical effects can be controlled \cite{salerno2022loss}. 
Adding checks on the symmetries should not be seen as an additional complication, since they can contribute largely to verifying the computation as physically correct. 

\vspace{0.25cm}
\textbf{Reciprocity}\\
For the linear interaction of matter and light, one can derive the following expression (Eq.~36 in \cite{mishchenko1996t}):
\begin{equation}
    T^{ij}_{l,m,l',m'} = (-1)^{m+m'} T^{ji}_{l',-m', l,-m}
\end{equation} 
for $l \in \mathbb N_0$, $m \in \mathbb Z$ with $l \geq |m|$, $i$ and $j$ the polarization indices, when the response of the object meets Lorentz reciprocity.
    
\vspace{0.25cm}
\textbf{Rotational invariance}\\
A rotationally symmetric object produces a scattering response that is also rotationally invariant. This sets constraints on the T-matrix of such an object, formulated in the following for the case of the symmetry axis aligned with the z-axis (Eqs.~30 and 31 in \cite{mishchenko1996t}):
\begin{align}
    T^{ij}_{l,m,l',m'} &= \delta_{mm'} T^{ij}_{l,m,l',m'}\, , 
\\
    T^{ij}_{l,m,l',m'} &= i j T^{ij}_{l,-m,l',-m'} \, .    
\end{align}
For a sphere, this condition implies a matrix with only diagonal entries in parity basis, which does not depend on azimuthal indices $m$, $m'$. Please note the difference in notation compared to the reference, as the polarization indices $i$, $j$ take the values of -1, 1.
    
\vspace{0.25cm}
\textbf{Mirror symmetries}\\
The T-matrix of an object invariant with respect to mirror symmetry operations exhibits certain constraints. The specific formula depends on the polarization and the plane of symmetry. As an example, for mirror reflection regarding the xy-plane, we can derive the formulas based on the properties of the spherical harmonics:
\begin{equation}
    Y_{lm}(\pi - \theta, \phi) = (-1)^{l+m} Y_{lm} (\theta, \phi)\, .
\end{equation}
For the parity basis, the following formula holds:
\begin{equation}
	T^{ij}_{l,m,l',m'} =  ij(-1)^{m + m' + l + l'} \, T^{ij}_{l,m,l',m'}\, ,
\end{equation}
In the helicity basis, the modified formula is:
\begin{equation}
	T^{ij}_{l,m,l',m'} =   (-1)^{m + m' + l + l'} \, T^{-i -j}_{l,m,l',m'}\, .
\end{equation}

\vspace{0.25cm}
\textbf{Lossless/non-absorptive}\\
For a non-absorbing object, the condition for the corresponding T-matrix can be derived by applying the principle of conservation of energy (Eq.~45 in \cite{mishchenko1996t}, Eq.~5.59 in~\cite{MishchenkoTL02}). The integral of the Poynting vector $\langle \mathbf{S(r)} \rangle \cdot \mathbf{r}$ over a spherical surface at infinity has to vanish (Eq.~5.55 in ~\cite{MishchenkoTL02}). This leads to the unitarity of the S-matrix: $ \mathbf{S}^{\dagger}\mathbf{S}=\mathbb{1}$. The corresponding T-matrix representation is then:
\begin{equation}\label{lossless}
    \mathbf{T}^{\dagger}\mathbf{T} = -\frac{1}{2}(\mathbf{T}^{\dagger} + \mathbf{T})\, , 
\end{equation} 
where $(\mathbf{T}^{\dagger})^{ij}_{l,m,l',m'} = 
(T^{ji}_{l',m',l,m})^*$. 

\vspace{0.25cm}
\textbf{Passive}\\
If loss is prevalent in the medium, the integral of the Poynting vector over
a spherical surface at infinity is negative, as more energy comes in than
goes out. From this inequality, it can be concluded that  the following expression is Hermitian positive-semidefinite~(Eq.~10 in~\cite{le2013radiative}):
\begin{equation}\label{hpsd}
    -2 \mathbf{T}^{\dagger}\mathbf{T} -\mathbf{T}^{\dagger} - \mathbf{T}\, .
\end{equation}
For the gain medium, the sign of the inequality is the opposite, thus the condition for positive-semidefineteness is violated.

\vspace{0.25cm}
\textbf{Numerical accuracy metric}\\
Since the T-matrix is obtained from the numerical solution of Maxwell's equations, the relations listed above are fulfilled with some degree of inaccuracy. It is not viable to demand these relations to hold exactly, but the discrepancy necessitates a single computable quantity. The following metric is introduced, however, this choice is not unique. For the metric to represent a relative deviation from the magnitude of the initial value, we take the sum of the squared absolute values of the difference between the transformed and initial matrix and divide it by the sum of the squared norm of all the initial and transformed matrix elements:
\begin{equation}
    \sigma =  \frac{1}{2} \frac{\sum_{l=1}^{l_{\text{max}}}  \sum_{l'=1}^{l_{\text{max}}} \sum_{m = -l}^{l} \sum_{m' = -l'}^{l'}  \sum_{i= \pm 1}  \sum_{j= \pm 1}{\lvert T^{ij}_{l,m,l',m'} - (T')^{ij}_{l,m,l',m'}} \rvert ^2}{\sum_{l=1}^{l_{\text{max}}}  \sum_{l'=1}^{l_{\text{max}}} \sum_{m = -l}^{l} \sum_{m' = -l'}^{l'}  \sum_{i= \pm 1}  \sum_{j= \pm 1}\lvert {T^{ij}_{l,m,l',m'}\rvert}^2 +  \lvert {(T')^{ij}_{l,m,l',m'}\rvert}^2}\, .
\end{equation}
This can be regarded as normalization by the total interaction cross-section, which is proportional to the scattering cross-section of the object. 
It is straightforward to apply the metric for geometric transformations. 
In case of Eq.~\ref{lossless}, the metric can be used with the transformed T-matrix reformulated as $\mathbf{T'} = 2\mathbf{T}^{\dagger}\mathbf{T} +\mathbf{T}^{\dagger}$. The Hermitian positive-semidefiniteness in Eq.~\ref{hpsd} is not tested using the metric. The minimum tolerance in Cholesky decomposition which produces the eigenvalues with non-negative values gives insight into the degree to which the condition in Eq.~\ref{hpsd} is fulfilled.   

\vspace{0.25cm}
\textbf{Convergence}\\
Since the goal is to make previously computed results reusable, additional checks on reliability are recommended. 
A convergence check is particularly important since a coarse mesh would not correctly represent the symmetries of the object, however, the algorithm of creating the mesh might also contribute to the emerging asymmetries. It is recommended for the researcher to check the mesh separately for the presence of symmetries existing in the object. No direct checks on the mesh are incorporated by default. 

An additional convergence study justifying the truncation at the chosen number of the multipole orders is highly valuable as well. For a single provided T-matrix, the average extinction cross-section can be calculated and compared with the same matrix truncated by one multipole order, while it is on the behalf of the researcher to decide whether a higher multipole order is needed. If the inclusion of the higher multipole order does not change the average extinction cross-section by more than one percent, this can be considered sufficient. One has, however, to keep in mind that for some more sensitive optical characteristics, relying on correct computation in the near-field region, the insufficiency of the number of multipole orders could be revealed.

The same approach is suggested for the other parameters. If data with different values of, e.g., resolution are calculated with admissible accuracy, it is still recommended to upload both cases and let the user decide if the accuracy is acceptable for their application. The provision of additional data ensuring convergence for all method parameters is not mandatory, and it is the responsibility of the researcher to ensure the trustworthiness of the data.

\part*{Examples of data usage}

This part is intended to describe examples of how researchers in the community can benefit from using the existing \texttt{.tmat.h5} files. It highlights the features of some of the software developed and indicates how to load and use files in the proposed data format. This should serve as a practical demonstration on how to capitalize on the data format. We consider the retrieval of T-matrices of two different objects and computation of their optical quantities with two different multiscattering tools. The use of these programs is only exemplary, and we encourage the readers to assess whether it is possible to add similar functions to their software to simplify the exchange of T-matrices.

\section*{TiO$_2$ cylinder}
\label{sec:treams}

We demonstrate a basic use case for the T-matrix data format. We have computed the T-matrix of a titanium dioxide cylinder, stored it in the suggested format, and finally we use a separate program to load the file and compute the ensemble-averaged scattering cross-section. The T-matrix is computed with JCMsuite as outlined in \cref{ch:jcm}. For the scattering cross-section calculation, we use \textit{treams}, an open-source T-matrix-based scattering program in Python \cite{beutel2024treams}. A module to load and store files, which conforms to the format described here, is part of the program.

\begin{figure}
    \centering
    \includegraphics[width=\linewidth]{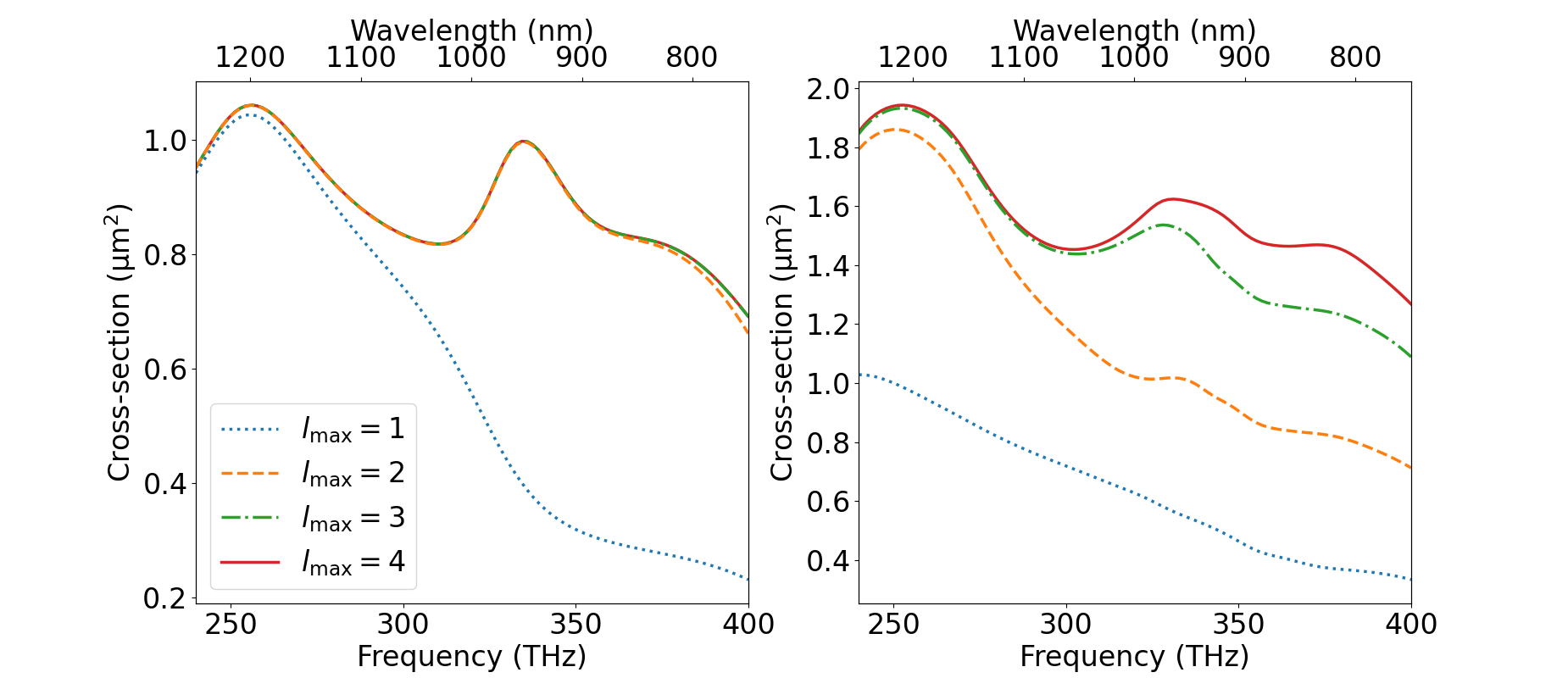}
    \caption{Average extinction cross-section of a single titanium dioxide cylinder (left) with radius \SI{250}{\nano\meter} and height \SI{300}{\nano\meter} and two of those cylinders separated horizontally by \SI{600}{\nano\meter} (right). }
    \label{fig:treams}
\end{figure}

In this example, we consider a cylinder made from titanium dioxide. The cylinder has a radius of \SI{250}{\nano\meter} and a height of \SI{300}{\nano\meter}. We consider a non-dispersive relative permittivity of $\epsilon = 6.25$ in the frequency range between \SI{240}{\tera\hertz} and \SI{400}{\tera\hertz}. We compute the T-matrix up to $l_\text{max} = 4$ using JCMsuite as outlined above. The main modification is the change of the layout file for the geometry of a cylinder. The full files can be found in Listing~\ref{lst:treams:full}. The created T-matrix file can then be loaded in \textit{treams} by \texttt{treams.io.load\_hdf5} as shown in the snippet below. The stored T-matrix is three-dimensional with the outer dimension corresponding to the number of frequencies.     
\begin{lstlisting}[language=python,caption=Data format usage in \textit{treams},firstline=11,lastline=26]
cyl = treams.io.load_hdf5("cylinder_tio2.tmat.h5")

positions = [[-300, 0, 0], [300, 0, 0]]
cyl_cluster = [
    treams.TMatrix.cluster([tm, tm], positions) for tm in cyl
]
cyl_cluster = [
    tm.interaction.solve().expand(
        treams.SphericalWaveBasis.default(4)
    )
    for tm in cyl_cluster
]

with h5py.File("two_cylinders.tmat.h5", "w") as fobj:
    treams.io.save_hdf5(fobj, cyl_cluster)
\end{lstlisting}

The T-matrix can then be directly used in \textit{treams}. In the example, we calculate the coupling between two disks separated horizontally by \SI{600}{\nano\meter}. The \texttt{cluster} method stacks the matrices together, and the \texttt{solve} method calculates the interaction between objects and outputs a T-matrix in the local coordinate systems of each object. This is then expanded at a new single origin to obtain a global T-matrix. Without going into further details of the functions used to calculate the interaction, we can store the final result equally easily by \texttt{treams.io.store\_hdf5}. The resulting average extinction cross-sections for the individual cylinder and the two cylinders are shown in Fig.~\ref{fig:treams}.

The main demonstrated feature is that the loading function automatically reads all the files and returns them as a list of T-matrices. The order of the entries with the multipole description and polarization, the distinction of parity and helicity basis, and the frequency are recognized and stored in the T-matrix class of \textit{treams}. Then, it can be conveniently used within \textit{treams}. The storing function converts the class objects for the T-matrix to the correct entries in the data file. Further information can then be added afterward, e.g., the description. The original data file is also available at~\cite{data_format}.

\section*{Gold spheroid}
In this example, we are interested in the T-matrix of a gold spheroid in water and consider a different multiscattering tool to exploit it. TERMS~\cite{schebarchov2022multiple, Schebarchov:2021ut} is a Fortran program based on the superposition T-matrix method, designed to simulate the near-field and far-field optical properties of collections of particles. It was developed primarily to model relatively compact clusters of resonant scatterers, such as plasmonic particles, often requiring large multipolar orders~\cite{schebarchov2019mind}. TERMS implements several independent algorithms, with complementary strengths and weaknesses, to describe the self-consistent electromagnetic interaction between multiple scatterers and compute far-field optical properties such as absorption, scattering, extinction, circular dichroism, as well as near-field intensities and the local degree of optical chirality. By describing the incident and scattered fields in a basis of spherical waves, the T-matrix framework lends itself to analytical formulas for orientation-averaged quantities such as far-field cross-sections and near-field quantities, greatly reducing the computational time needed to simulate particles and systems of particles in random orientation~\cite{fazel2022orientation}. Each scatterer is described by a T-matrix, which is computed internally for spherical particles (including layered spheres), or using external files computed with any other method.

The program's documentation and website~\cite{Schebarchov:2021ut} offer many examples of use; here we only illustrate the import of an external T-matrix in the \texttt{tmat.h5} format. The input file for the simulation reproduced below considers two gold spheroids in water, separated by 100~nm and rotated by 45 degrees to form a chiral structure.

\begin{lstlisting}[label={lst-terms},language=Fortran, caption=Script for TERMS]
ModeAndScheme 2 3
MultipoleCutoff 3 3 -3
Wavelength 400 800 200
Medium 1.7689
OutputFormat HDF5 smarties_dimer
TmatrixFiles 1 
smarties.tmat.h5
Scatterers 2
TF1 0 -50 0 40 0 0 0 2
TF1 0  50 0 40 0 0.7853982 0 2
\end{lstlisting}





\begin{figure}[h!]
\centering{
\includegraphics[width=0.9\linewidth]{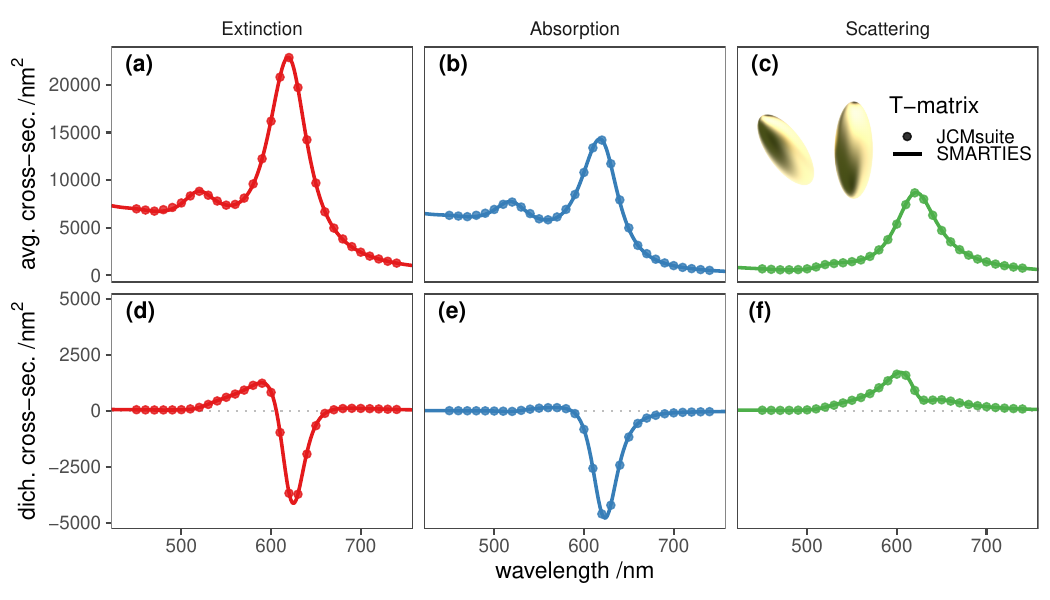}
}
\caption{\label{fig-spheroid-dimer}Validation of results obtained using a T-matrix computed with JCMsuite (points), and SMARTIES (solid lines), both imported into TERMS for a multiple scattering calculation (Listing~\ref{lst-terms}). The structure consists of a dimer of identical Au spheroids immersed in water, with center-to-center separation 100~nm, and a dihedral angle of \(\pi/4\) making the dimer chiral. The top panels (a--c) present the orientation-averaged optical cross-sections for extinction, absorption, and scattering, respectively. The bottom panels (d--f) present the corresponding orientation-averaged circular dichroism cross-sections. Excellent agreement is obtained between the two methods.}
\end{figure}

The simulation is run with the command \texttt{terms\ input}, and outputs cross-sections in the file \texttt{results.h5}. The results are displayed in Fig.~\ref{fig-spheroid-dimer}. For comparison, the same simulation was run with a T-matrix produced by SMARTIES (Listing~\ref{lst-smarties}) for the same geometry.

Since Fortran is a low-level language, it is not very practical to implement a full support of all the options of the \texttt{tmat.h5} format. Instead, TERMS has currently implemented a basic import functionality with following expectations. The \texttt{vacuum\_wavelength} is the only allowed field, and must be provided in nanometers. The wavelengths must match exactly the TERMS input file. Furthermore, no check is performed for the relative permittivity of the embedding medium, which should match the TERMS input file.

Hopefully, a Python interface to TERMS will be available in the future, which will add flexibility in the import of external T-matrices.

\part*{Summarizing statements}
\section{Summary}

This manuscript establishes a unified data format for storing and distributing T-matrices, which represents a pivotal step in handling scattering data within the scientific community. T-matrices, in general, contain the complete information of the scattering properties of a scatterer in linear approximation. They tell us how light, or, more generally, electromagnetic waves, interact with a given object. However, they can also be used to express the properties of more advanced photonic materials made from many scatterers. Adopting this standardized format is a critical step toward addressing several prevalent challenges faced by researchers, including the repetitive computation of T-matrices across many laboratories worldwide. These computations frequently require considerable computational resources and pose significant financial and environmental burdens.

By providing a common framework for the systematic storage, accessibility, and sharing of T-matrices, we enhance the ability to reuse previously calculated data, thereby significantly reducing the need for duplicate computations. This approach conserves valuable resources and lessens the ecological impact of high-performance computing activities, which is increasingly important in light of current global energy concerns. Furthermore, the standardization of T-matrix data ensures that the scientific results are reproducible and verifiable, which is essential for maintaining the integrity and reliability of photonic research. 

To permit wide use of this data format, we have described in this contribution multiple approaches to computing T-matrices and storing them in the required data format. In addition to the description provided here, the supporting files that can be used by others to compute T-matrices and store them in the expected data format are potentially of more practical use. These files are publicly available at~\cite{data_format}.

We wish to motivate further scientists to adapt their tools so that T-matrices can be generated and stored in the desired format. The more tools are verified for this purpose, the larger the number of scientists that can benefit from the standardized manner of storing and archiving T-matrices. Moreover, we also demonstrated how to use T-matrices in a set of programs besides generating T-matrices. Along the same lines, we wish to motivate many more scientists to seek ways to integrate these T-matrices into their computational workflows.

The proposed data format is designed to be flexible yet precise enough to accommodate T-matrices of a wide range of different scattering structures, reflecting the needs of various spectral domains. This flexibility ensures broad applicability across multiple disciplines, including optics and photonics, nanotechnology, environmental sciences, microwave scattering, and biotechnology, among others. As such, researchers can now access a wealth of T-matrix data that may previously have been recalculated redundantly, accelerating the pace of innovation in fields reliant on scattering data.

Moreover, the uniform data format facilitates the integration of T-matrices with emerging technologies and methodologies, such as machine learning and big data analytics, which are becoming increasingly prevalent in scientific research. By enabling the efficient use of T-matrices in training machine learning-based technology to solve direct and inverse scattering problems, this format could potentially prevent the need for direct numerical simulations in some cases, streamlining the research process significantly \cite{wiecha2021deep,nees2023multi, gladyshev2023inverse,khaireh2023newcomer}. This increasingly large number of simulations conducted solely with the purpose of generating training data, which is frequently computationally expensive, would no longer be necessary if systematically generated data from many labs worldwide could be used for training purposes.

The impact of this standardization extends beyond simplification and cost reduction in computational processes. It also promotes a collaborative scientific environment where researchers across the globe can contribute to and benefit from a future shared repository of T-matrices. This collaborative approach fosters innovation and enhances the educational value of scattering data, allowing for more comprehensive and advanced training of future scientists. To this end, it significantly supports research and development in optical nanometrology and advanced applications, ranging from nanotechnologies and advanced (nano-) materials to novel photonic devices. 

\section{Outlook}

This establishment of a standardized data format for T-matrices represents only the initial phase of a longer initiative aimed at enhancing the accessibility and utility of scattering data within the scientific community. The next crucial step in this endeavor is the development of an online database that facilitates not only the uploading but also the sharing of T-matrix data. This proposed database will be designed with advanced search capabilities, ideally leveraging large language models, allowing users to efficiently retrieve T-matrices that closely resemble a specific object or set of scattering properties. Additionally, it will enable users to input a desired T-matrix and search for corresponding entries in the database that match the input at a predefined operational frequency. This functionality is particularly valuable as it effectively addresses the inverse problem, offering a powerful tool for researchers seeking specific scattering responses without the need for direct computation. Such a resource will significantly expedite the process of scientific discovery and innovation, providing a robust platform for researchers worldwide to collaborate and advance the field of photonics and related disciplines.

Establishing an online and open access T-matrix database raises several research questions and opportunities, particularly in comparing the accuracy and efficiency of various computational methods used to generate T-matrices. Researchers could systematically evaluate the performance of different scattering computation techniques, such as the finite element method, boundary element method, finite difference time domain method, or discrete dipole approximation, by comparing their output T-matrices stored within the database. This analysis would reveal discrepancies, validate the methods across various materials and geometries, and help refine these computational techniques for enhanced accuracy and efficiency.

Furthermore, the database opens the door to address interesting research questions. By mining the database for unique T-matrix patterns and associating them with physical structures, researchers could identify scatterers with predefined properties. These explorations could lead to the discovery of new materials with applications in advanced photonic technologies like lighting devices, enhanced sensing, or light management in solar cells.

Finally, this database democratizes access to high-level photonic research, allowing researchers without the computational expertise or resources for the computation of T-matrices to participate actively in the field. The database adds diversity to the research community by providing pre-computed T-matrices and enhances collaborative opportunities across different domains. This inclusive approach could lead to new perspectives and innovative uses of photonic systems.

\section{Conclusions}

In conclusion, establishing a standard data format for T-matrices is a significant step in our scientific community's efforts to optimize research efficiency and collaboration. It addresses economic and ecological issues associated with repetitive computations and paves the way for novel research opportunities in photonics and related fields. The scientific community can look forward to more sustainable, reproducible, and innovative research outcomes by adhering to this standardized approach. The future database, where the datasets can be shared, in contrast to a \enquote{data lake} with unstructured data, serves as an ideal platform for finding use cases and datasets for training and testing out machine learning methods in the field of photonics. This is an area which is up to now underserved in the classical data repositories for machine learning such as Kaggle, opening a unique opportunity to establish a go-to place for popularizing otherwise niche research data. The ongoing development and widespread adoption of this data format will undoubtedly catalyze further advancements in studying and applying complex photonic systems, with broad implications across multiple scientific disciplines.

\section*{Acknowledgements}
N.A., K.B., J.M., and C.R. acknowledge support by the Federal Ministry of Education and Research (BMBF) within the project DAPHONA (16DKWN039).
K. Achouri acknowledges funding from the Swiss National Science Foundation (Grant No. TMSGI2\_218392).
B.A. and E.C.L.R. thank the Royal Society of New Zealand Te Ap\=arangi for support through Marsden grants MFP-VUW2204 and MFP-VUW2118. 
D.B., F.T., and C.R. acknowledge support by the Deutsche Forschungsgemeinschaft (DFG, German Research Foundation) under Germany's Excellence Strategy via the Excellence Cluster 3D Matter Made to Order (Grant No. EXC - 2082/1 - 390761711) and from the Carl Zeiss Foundation via CZF-Focus@HEiKA. 
K.M.C. acknowledges support by the Polish National Science Center via the project
2020/37/N/ST3/03334.
S.B. acknowledges support by BMBF (Forschungscampus  MODAL, project number 05M20ZBM) and by DFG under Germany's Excellence Strategy - The Berlin Mathematics Research Center MATH+ (EXC-2046/1, project ID: 390685689).
J.M. acknowledges the Presidential Sejong fellowship (RS-2023-00252778) funded by the Ministry of Science and ICT (MSIT) of the Korean government. 
H.K. acknowledges the Asan Biomedical Science fellowship, and the Presidential Science fellowship funded by the MSIT of the Korean government. 
J.R. acknowledges the POSCO-POSTECH-RIST Convergence Research Center program funded by POSCO, and the NRF grant (RS-2024-00356928) funded by the MSIT of the Korean government.
A.B. acknowledges support from the French National Research Agency (ANR) in the frame of the project MELODIE (Grant ANR-22-CE09-0027) and by the Institut Universitaire de France (IUF).
D.G. and S.R. acknowledge support by the Austrian Science Fund (FWF) under project P32300 (WAVELAND). 
M.Y. acknowledges support of the Normandy Region (project RADDAERO).
L.P. acknowledges support by the European project 21GRD03 PaRaMetriC. The project 21GRD03 PaRaMetriC received funding from the European Partnership on Metrology, co-financed by the European Union's Horizon Europe Research and Innovation Programme and from the Participating States.
K. Arjas and P.T. acknowledge support from the Academy of Finland under Project No. 349313 and from the Vilho, Yrj\"o and Kalle V\"ais\"al\"a Foundation.
A.B.E. acknowledges the support from the Deutsche Forschungsgemeinschaft (DFG, German Research Foundation) under Germany's Excellence Strategy within the Cluster of Excellence PhoenixD (EXC 2122, Project ID 390833453).
B.B. acknowledges support by the European project 20FUN02 ``POLight''. The work has been partly addressed also in the project 20FUN02 ``POLight''. The project 20FUN02 ``POLight'' has received funding from the EMPIR programme co-financed by the Participating States and from the European Union's Horizon 2020 research and innovation programme.
L.P. and K.C. acknowledge fruitful collaboration with the main developer of \textsc{smuthi} and \textsc{celes}, Amos Egel. 
L.P. acknowledges funding by the German Research Foundation-Project-ID 416229255-SFB 1411.
R.V. acknowledges Guillaume Dem{\'e}sy (Institut Fresnel, Marseille) for interactions on the use of the free ONELAB FEM software to calculate the T-matrix of peculiar structures.
We thank Pascal Scherer that implemented the codes to use MEEP in the calculation of the T-matrix. 

\newpage

\appendix

\part*{Appendix}\addcontentsline{toc}{part}{Appendix}

\section{Constitutive relations}
\label{app:constitutiverelations}

We consider causal materials invariant under translations of time, such that a description by time-harmonic fields with a fixed frequency can be used (please be reminded that we use the $\exp(-\mathrm \mathrm{i} \omega t)$ convention here). Thus, in principle, all quantities in this chapter are defined for dispersive media. However, the frequency argument will be omitted here. The most general case of constitutive relations for linear homogeneous materials are
\begin{equation}
    \begin{pmatrix}
        \frac{1}{\epsilon_0}\mathbf{D} \\
        c_0\mathbf{B}
    \end{pmatrix}
    =
    \begin{pmatrix}
        \boldsymbol{\epsilon} & \boldsymbol{\xi} \\
        \boldsymbol{\zeta} & \boldsymbol{\mu}
    \end{pmatrix}
    \begin{pmatrix}
        \mathbf{E} \\
        Z_0\mathbf{H}
    \end{pmatrix}
\end{equation}
with dimensionless 3-by-3 tensors $\boldsymbol{\epsilon}$, $\boldsymbol{\mu}$,
$\boldsymbol{\xi}$, and $\boldsymbol{\zeta}$. We refer to the full 6-by-6 tensor as bianisotropic tensor. The quantities $\epsilon_0$, $c_0$, and $Z_0$ are the vacuum values of the permittivity, speed of light, and the wave impedance, respectively. These prefactors are chosen to normalize all fields to the same unit, namely \si{\volt\per\meter}. Thus, the bianisotropic tensor contains dimensionless units. If all four 3-by-3 tensors are proportional to the unit matrix, then the material is called biisotropic. Then, the material parameters can be expressed as scalars $\epsilon$, $\mu$, $\xi$, and $\zeta$. The magnetoelectric couplings can be expressed alternatively with the non-reciprocity parameter $\chi = \tfrac{\xi + \zeta}{2}$ and the chirality parameter $\kappa = \tfrac{\xi - \zeta}{2 \mathrm{i}}$ as
\begin{align}
    \begin{pmatrix}
        \frac{1}{\epsilon_0}\mathbf{D} \\
        c_0\mathbf{B}
    \end{pmatrix}
    &=
    \begin{pmatrix}
        \epsilon & \xi \\
        \zeta & \mu
    \end{pmatrix}
    \begin{pmatrix}
        \mathbf{E} \\
        Z_0\mathbf{H}
    \end{pmatrix}
    \\
    \label{eq:biisotropic}
    &=
    \begin{pmatrix}
        \epsilon & \chi + \mathrm{i} \kappa \\
        \chi - \mathrm{i} \kappa & \mu
    \end{pmatrix}
    \begin{pmatrix}
        \mathbf{E} \\
        Z_0\mathbf{H}
    \end{pmatrix}
    \,.
\end{align}
If the non-reciprocity parameter vanishes ($\chi = 0$), then the material is referred to as chiral.

Starting from the bianisotropic case, but requiring vanishing magnetoelectric coupling, i.e., $\boldsymbol{\xi} = 0 = \boldsymbol{\zeta}$, instead of isotropy, leads to the constitutive relations
\begin{equation}
    \begin{pmatrix}
        \frac{1}{\epsilon_0}\mathbf{D} \\
        c_0\mathbf{B}
    \end{pmatrix}
    =
    \begin{pmatrix}
        \boldsymbol{\epsilon} & 0 \\
        0 & \boldsymbol{\mu}
    \end{pmatrix}
    \begin{pmatrix}
        \mathbf{E} \\
        Z_0\mathbf{H}
    \end{pmatrix}
\end{equation}
of an anisotropic material. Finally, if the material is isotropic and has no magnetoelectric coupling, then we have an isotropic material with
\begin{equation}
    \begin{pmatrix}
        \frac{1}{\epsilon_0}\mathbf{D} \\
        c_0\mathbf{B}
    \end{pmatrix}
    =
    \begin{pmatrix}
        \epsilon & 0 \\
        0 & \mu
    \end{pmatrix}
    \begin{pmatrix}
        \mathbf{E} \\
        Z_0\mathbf{H}
    \end{pmatrix}
\end{equation}
as constitutive relations.
In a general case, the material can be treated as nonlocal, which implies that the electric field at a point $\mathbf{r}$ is influenced not only by the electric field at that point, but also at all other points $\mathbf{r^{\prime}}$ within a spatial domain surrounding $\mathbf{r}$. The following relation holds:
\begin{equation}
    \mathbf{D}(\mathbf{r}) = \int \mathbf{R}(\mathbf{r}-\mathbf{r^{\prime}}) \mathbf{E(r^{\prime})} \mathrm{d}{\mathbf{r^{\prime}}}\, ,
    \end{equation}
 where $\mathbf{R}(\mathbf{r}-\mathbf{r^{\prime}})$ is the  nonlocal response kernel. In~\cite{Doicu2020}, the nonlocal response equation is derived in the following form:
\begin{equation}
    \left( \frac{\beta^2}{\omega^2 + \mathrm{i} \gamma \omega} - \mathrm{i} \frac{D}{\omega} \right) \nabla \cdot (\nabla \cdot \mathbf{J} (\mathbf{r}))  + \mathbf{J} (\mathbf{r}) = \sigma \mathbf{E} (\mathbf{r})
\end{equation}    
where
\begin{equation}
\sigma = \mathrm{i}\epsilon_0 \frac{\omega_\text{p}^2}{\omega + \mathrm{i}\gamma}
\end{equation}
is the Drude conductivity,
\begin{equation}
\omega_\text{p}^2 = \frac{n_0 e^2}{\epsilon_0 m}
\end{equation}
the plasma frequency of the metal, $\gamma$ is the Drude damping rate, $n_0$ the equilibrium electron density, $\beta^2 = (3/5)v_\text{F}^2$, here $v_\text{F}$ is the Fermi velocity, $D$ is the diffusion constant, and $e$ and $m$ are the  electron charge and mass, respectively.

\section{Coordinate systems}
\label{app:coordinates}
The Cartesian, cylindrical, and spherical coordinates are related by
\begin{equation}
    \label{eq:coordinates}
    \begin{pmatrix}
        x \\ y \\ z
    \end{pmatrix}
    =
    \begin{pmatrix}
        \rho \cos{\phi} \\
        \rho \sin{\phi} \\
        z
    \end{pmatrix}
    =
    \begin{pmatrix}
        r \sin{\theta} \cos{\phi} \\
        r \sin{\theta} \sin{\phi} \\
        r \cos{\theta}
    \end{pmatrix}
\end{equation}
and have the associated unit vectors
\begin{align}
    \label{eq:unitvectors}
    \begin{pmatrix}
        \boldsymbol{\hat{r}} \\
        \boldsymbol{\hat{\theta}} \\
        \boldsymbol{\hat{\phi}}
    \end{pmatrix}
    &=
    \begin{pmatrix}
        \sin{\theta} & 0 & \cos{\theta} \\
        \cos{\theta} & 0 & -\sin{\theta} \\
        0 & 1 & 0
    \end{pmatrix}
    \begin{pmatrix}
        \boldsymbol{\hat{\rho}} \\
       \boldsymbol{\hat{\phi}} \\
       \boldsymbol{\hat{z}}
    \end{pmatrix}
    \\
    &=
    \begin{pmatrix}
        \sin{\theta} \cos{\phi} & \sin{\theta} \sin{\phi} & \cos{\theta} \\
        \cos{\theta} \cos{\phi} & \cos{\theta} \sin{\phi} & -\sin{\theta} \\
        -\sin{\phi} & \cos{\phi} & 0
    \end{pmatrix}
    \begin{pmatrix}
        \mathbf{\hat{x}} \\
        \mathbf{\hat{y}} \\
        \mathbf{\hat{z}}
    \end{pmatrix}
    \,.
\end{align}
By default, the z-axis has a special role (symmetry axis of an axisymmetric object). If this is changed to either the x or y-axis, $(x, y, z)^T$ in \Cref{eq:coordinates} is replaced by $(y, z, x)^T$ or $(z, x, y)^T$, respectively. \Cref{eq:unitvectors} is changed accordingly to $(\mathbf{\hat{y}}, \mathbf{\hat{z}}, \mathbf{\hat{x}})^T$ or $(\mathbf{\hat{z}}, \mathbf{\hat{x}}, \mathbf{\hat{y}})^T$.

\section{Mode normalization}
\label{app:normalization}

Here, we comprehensively define the normalization of the modes starting from elementary functions, which coincides with the normalization in \cite{jackson1998classical}. This is necessary to have an unambiguous definition of the vector spherical waves we use. We also define a reference T-matrix of an object which can be used to verify the normalization. Please note, we work here with complex-valued VSWF. If real-valued expressions are needed, dedicated conversion tools can easily be set up.  

\subsection{Definition}

We start with the definition of the associated Legendre polynomials (which are, in general, no polynomials) by
\begin{equation}
    P_l^m(x)
    = \frac{(-1)^m}{2^l l!}
    (1 - x^2)^\frac{m}{2}
    \frac{\mathrm d^{l + m}}{\mathrm d x^{l + m}}
    (x^2 - 1)^l
\end{equation}
for $l \in \mathbb N_0$ and $m \in \mathbb Z$ with $l \geq |m|$. Please note especially the factor $(-1)^m$ in front. With the associated Legendre polynomials as above, we define the spherical harmonics by
\begin{equation}
    Y_{lm}(\theta, \phi)
    = \sqrt{\frac{2l + 1}{4 \pi} \frac{(l - m)!}{(l + m)!}}
    P_l^m(\cos\theta)
    \mathrm e^{ \mathrm{i} m \phi}\,.
\end{equation}
We define the vector spherical waves by:
\begin{equation}\label{Eq:C3}
    \mathbf{M}_{lm}^{(n)}(kr, \theta, \phi) = z_l^{(n)}(kr)   \mathbf{X}_{l, m}(\theta, \phi)\, .
\end{equation}
 Here, $ \mathbf{X}_{lm}$ is the normalized vector spherical harmonics:


\begin{align}
  &\mathbf{L} =   \frac{\mathbf{r} \times \nabla}{\mathrm{i}} \\
     & \mathbf{X}_{l, m}(\theta, \phi) = \frac{1}{\sqrt{l(l+1)}} \mathbf{L} Y_{l, m}(\theta, \phi)
\end{align}
The superscript $n = 1$ refers to the incident modes and $n = 3$ is used for the scattered modes. These are the modes that matter in the context of the T-matrix definition. 
Thus, the functions $z_l^{(1)}(x) = j_l(x)$ are the spherical Bessel functions, and $z_l^{(3)}(x) = h_l^{(1)}(x)$ are the spherical Hankel functions of the first kind. For completeness, the functions $z_l^{(2)}(x) = N_l(x)$ are the spherical Neumann functions, and $z_l^{(4)}(x) = h_l^{(2)}(x)$ are the spherical Hankel functions of the second kind.

For the time evolution, we use $\exp(-\mathrm i \omega t)$ such that the spherical Hankel functions of the first kind are traveling outwards. $k$ is the wavenumber in the medium. After applying the operator to the spherical harmonics, the vector spherical waves $\mathbf{M}_{lm}^{(n)}(kr, \theta, \phi)$ can be expressed as
\begin{equation}
    \begin{split}
        \mathbf{M}_{lm}^{(n)}(kr, \theta, \phi)
        =
        z_l^{(n)}
        \bigg[
            \frac{1}{2}
            \left(
                \lambda_+ Y_{l, m + 1}(\theta, \phi)
                + \lambda_- Y_{l, m - 1}(\theta, \phi)
            \right)
            \mathbf{\hat{x}}&
            \\
            +
            \frac{1}{2 \mathrm i}
            \left(
                \lambda_+ Y_{l, m + 1}(\theta, \phi)
                - \lambda_- Y_{l, m - 1}(\theta, \phi)
            \right)
            \mathbf{\hat{y}}&
            \\
            +
            m Y_{lm}(\theta, \phi) \mathbf{\hat{z}}&
        \bigg]
    \end{split}
\end{equation}
with $\lambda_\pm =\frac{\sqrt{(l \mp m)(l \pm m + 1)}}{\sqrt{l(l+1)}}$ or equivalently as
\begin{align}
    \mathbf{M}&_{lm}^{(n)}(kr, \theta, \phi)
    \notag \\
    &=
    \mathrm i \sqrt{\frac{(2l + 1)}{4 \pi l (l + 1)} \frac{(l - m)!}{(l + m)!}}
    \bigg[
        \mathrm i
        \frac{m P_l^m(\cos\theta)}{\sin\theta}
        \boldsymbol{\hat{\theta}}
        -
        \frac{\partial}{\partial \theta}
        P_l^m(\cos\theta)
        \boldsymbol{\hat{\phi}}        
    \bigg]
    \mathrm e^{\mathrm{i} m \phi}
    z_l^{(n)}(kr)\,.
\end{align}
This mode used for the electric field is called \enquote{TE} (transverse electric) mode or magnetic multipole. An orthogonal mode to this one can be defined by $\mathbf{N}_{lm}^{(n)}(kr, \theta, \phi) = \frac{\nabla}{k} \times \mathbf{M}_{lm}^{(n)}(kr, \theta, \phi)$ which results in
\begin{align}
    \mathbf{N}&_{lm}^{(n)}(kr, \theta, \phi)
    \notag \\
    &=
    \mathrm i \sqrt{\frac{(2l + 1)}{4 \pi l (l + 1)} \frac{(l - m)!}{(l + m)!}}
    \bigg[
        \left(
            \frac{\partial}{\partial \theta}
            P_l^m(\cos\theta)
            \boldsymbol{\hat{\theta}}
            +
            \mathrm i
            \frac{m P_l^m(\cos\theta)}{\sin\theta}
            \boldsymbol{\hat{\phi}}    
        \right)
        \frac{1}{k}
        \frac{\partial}{\partial r}
        (kr z_l^{(n)}(kr))
        \notag \\
        &\quad
        +
        l (l + 1)
        P_l^m(\cos\theta)
        \mathbf{\hat{r}}
        z_l^{(n)}(kr)
    \bigg]
    \frac{\mathrm e^{\mathrm{i} m \phi}}{kr}\, ,
\end{align}
which is used for the mode named \enquote{TM} (transverse magnetic) or electric multipole. 

Finally, we define the modes for positive and negative helicity by
\begin{equation}
    \mathbf{A}_{lm\pm}^{(n)}(k_\pm r, \theta, \phi)
    =
    \frac{
        \mathbf{N}_{lm}^{(n)}(k_\pm r, \theta, \phi)
        \pm \mathbf{M}_{lm}^{(n)}(k_\pm r, \theta, \phi)
    }{\sqrt{2}}\,.
\end{equation}
For biisotropic materials, only modes with well-defined helicity are solutions for Maxwell's equations with the constitutive relations from \Cref{eq:biisotropic}. The wavenumber
\begin{equation}
    k_\pm = k_0 \left(\sqrt{\epsilon \mu - \chi^2} \pm \kappa\right)
\end{equation}
becomes then polarization dependent.

\subsection{Reference T-matrix}
\label{app:reference}
We define a reference object as a verification aid for the T-matrix normalization. It consists of differently-sized spheres made from homogenous materials. It is chosen such that it does not have vanishing entries in the T-matrix, so all spatial symmetries are broken. The symmetry breaking is achieved by using differently sized spheres at the corners of a tetrahedron. All spheres are made of a material with relative permittivity $\epsilon = 9$. The embedding medium is vacuum, and the wavelength is in the range from $300$ to $\SI{700}{\nano\metre}$.
The spheres' radii and positions are: 

\begin{itemize}
    \item $r_1 = \SI{50}{\nano\metre}$ and
        $\mathbf{a}_1 = a
        \left(-\frac{1}{2}, -\frac{\sqrt{3}}{6}, -\frac{\sqrt{6}}{12}\right)$
    \item $r_2 = \SI{60}{\nano\metre}$ and
        $\mathbf{a}_2 = a
        \left(+\frac{1}{2}, -\frac{\sqrt{3}}{6}, -\frac{\sqrt{6}}{12}\right)$
    \item $r_3 = \SI{70}{\nano\metre}$ and
        $\mathbf{a}_3 = a \left(0, \frac{\sqrt{3}}{3}, -\frac{\sqrt{6}}{12}\right)$
    \item $r_4 = \SI{80}{\nano\metre}$ and
        $\mathbf{a}_4 = a \left(0, 0, \frac{\sqrt{6}}{4}\right)$
\end{itemize}
with the side length $a = \SI{300}{\nano\metre}$ of the tetrahedron. These values were selected arbitrarily.   

\begin{figure}[h!!]
  \centering
  \includegraphics[width=1\linewidth]{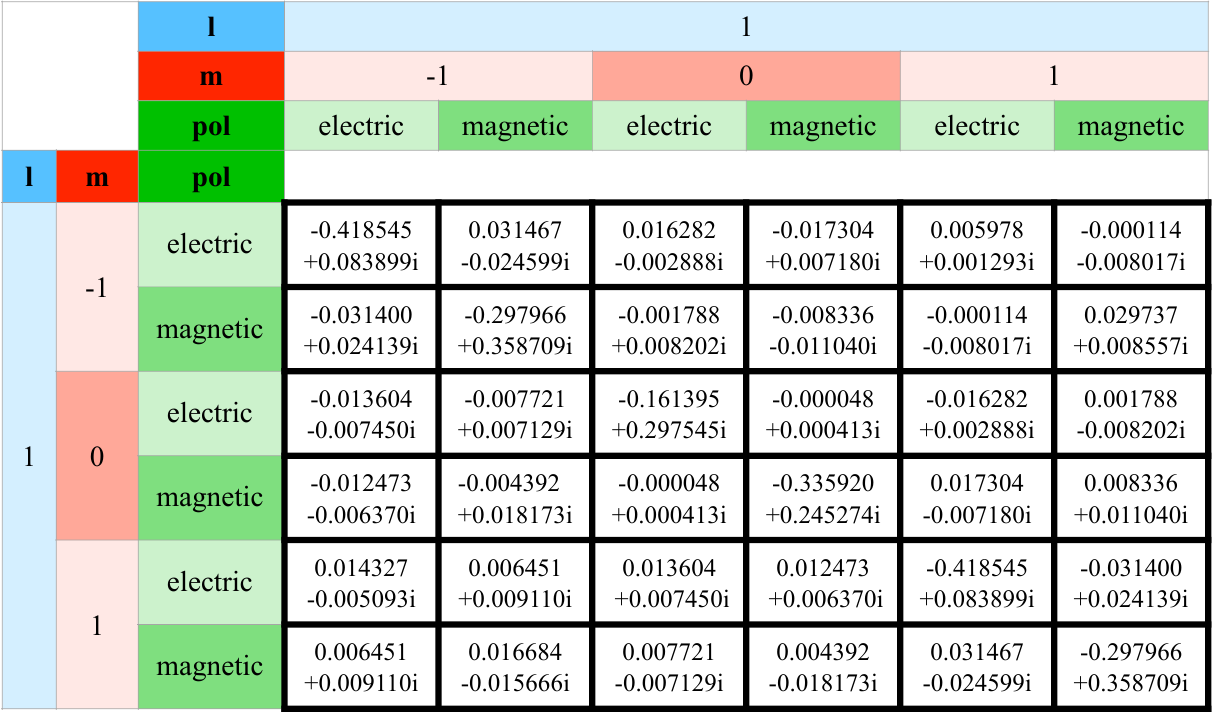}
  \caption{First 6$\times$6 entries of the global T-matrix for a reference arrangement of 4 spheres following the ordering described previously.}
  \label{fig:tabulated}
\end{figure}

The T-matrix of this structure was calculated semi-analytically using \emph{treams} software. Multipole order up to 6 was considered. At $\SI{500}{\nano\metre}$, the average extinction cross-section is \SI{ 0.2141779}{\micro \meter\squared}. The first 6$\times$6 entries of the global T-matrix in parity basis are presented in Fig.~\ref{fig:tabulated}. 

The provided HDF5 file contains all the necessary information and can be accessed via~\cite{data_format}.
\section{HDF5 file format and conversion tools}
\label{sec:hdf5}

First, this section gives a very brief introduction to the different components of the HDF5 file format and serves the purpose of explaining why it is a valuable framework for storing the considered data. We also emphasize particular aspects that need to be considered in setting up the data format. Additionally, it defines the general idea of how different pieces of data are related.
Next, some particular options for conversion of files from other formats are specified.

\subsection{HDF5 file format}
HDF5 is a hierarchic file format. Its main components are \emph{groups}, which work similarly to directories, and \emph{datasets}, which contain the actual data similar to files. However, more information about the data type is provided compared to regular files. These \emph{datasets} can be arrays of an arbitrary number of dimensions. It is possible to specify a dataset to have certain initial dimensions, and certain maximum size of the dataset, thus keeping an option to add more data to the dataset. Each item can have \emph{attributes} associated, which are small pieces of data directly attached to a \emph{group} or \emph{dataset}. Usually, they are used to providing metadata. Different \emph{groups} or \emph{datasets} can be linked with \emph{softlinks} or \emph{hardlinks}.

A complication in using HDF5 is the lack of a native complex number data type. Typically, complex numbers are stored as compound data types consisting of two floating-point numbers. The names of the fields are often \enquote{r} and \enquote{i}, but other conventions also exist. It is highly recommended to use the mentioned field names if possible.

Another issue with HDF5 is the order of the array, namely if row-major or column-major order is used. HDF5 itself uses row-major order~\cite{hdf5}. However, some programming languages use column-major order natively, like Fortran, MATLAB, or Julia. This can lead to issues when exchanging data between programs that use a different order, if the standard HDF5 Fortran wrapper is not used. To avoid ambiguity, the final formatting in row-major order is expected. For Python scripts, this is the default behavior, while for MATLAB, Julia, we refer to a supplementary link in~\cite{data_format}.

Finally, it should be mentioned that in the description of data entries we are following the type definitions in Python. As such, by claiming that type of the attribute is a string, we refer to \texttt{str} text type of a variable, which corresponds to \texttt{char} type in MATLAB. By claiming that a dataset is a one-dimensional array, a Numpy array is referred to, which does not have any additional dimensions of length 1. Thus, it is equivalent to a vector in MATLAB definitions. In some languages, the identical representations are not easily attainable, thus known alternatives in other languages will be considered and converted correspondingly when reading into the database.

\subsection{Tools for data format conversion}\label{tools-for-conversion}
Several open-source programs are available to compute T-matrices, but many of them do not (yet) implement the output format presented herein. While we encourage the community to add this functionality in order to fully benefit from interoperability between programs, it can also be useful, as a short-term or one-time workaround, to \emph{convert} T-matrix data stored in a different form. One example is the ``long format'' used to store T-matrix entries in earlier versions of SMARTIES~\cite{somerville2016smarties}, or Scuff-EM~\cite{Reid:vp}, or PXTAL~\cite{Garcia-de-Abajo:1999aa}, among others. We include example scripts at~\cite{data_format} to reshape such data and produce a standard \texttt{.h5} format.

Conversion to wide format, and export as \texttt{.tmat.h5}, can then be done by adding the required geometry and material information to make a complete entry. Basic export scripts are available in 4 different languages (R, Julia, MATLAB, Python) to serve as examples for similar conversion tasks.

\section{Units}
\label{app:units}
The accepted units are listed below, \enquote{Hz} can be substituted by \enquote{s\textasciicircum\{\nobreakdash-1\}}. For inverse length unit,\enquote{\textasciicircum\{\nobreakdash-1\}} should be appended to the length unit. 
\begin{table}[h!!]\label{tab:units}
  \centering
  \begin{tabular}{ | m{3cm} | m{3cm} | m{3cm} | }
  \hline
Value &  FREQUENCIES & LENGTHS \\
    \hline
   1e-24 & \enquote{yHz} & \enquote{ym}\\
    1e-21 & \enquote{zHz} &  \enquote{zm}\\
    1e-18 & \enquote{aHz} & \enquote{am}\\
    1e-15 & \enquote{fHz} & \enquote{fm}\\
    1e-12 & \enquote{pHz} & \enquote{pm}\\
    1e-9 & \enquote{nHz} & \enquote{nm}\\
    1e-6 & \enquote{uHz} & \enquote{um}\\
    1e-3 & \enquote{mHz} & \enquote{mm}\\
    1e-2 & \enquote{cHz} & \enquote{cm} \\
    1e-1 & \enquote{dHz} & \enquote{dm}\\
    1 & \enquote{Hz} & \enquote{m}\\
    1e1 & \enquote{daHz} & \enquote{dam}\\
    1e2 & \enquote{hHz} & \enquote{hm}\\
    1e3 & \enquote{kHz} & \enquote{km}\\
    1e6 & \enquote{MHz} & \enquote{Mm}\\
    1e9 & \enquote{GHz} & \enquote{Gm}\\
    1e12 & \enquote{THz} & \enquote{Tm}\\
    1e15 & \enquote{PHz} & \enquote{Pm}\\
    1e18 & \enquote{EHz} & \enquote{Em}\\
    1e21 & \enquote{ZHz} & \enquote{Zm} \\
    1e24 & \enquote{YHz} & \enquote{Ym}
    \\ \hline
  \end{tabular}
\end{table}
\newpage

\section{Example structures}
\begin{figure}[h!!]
\centering
\includegraphics[width=0.9\linewidth]{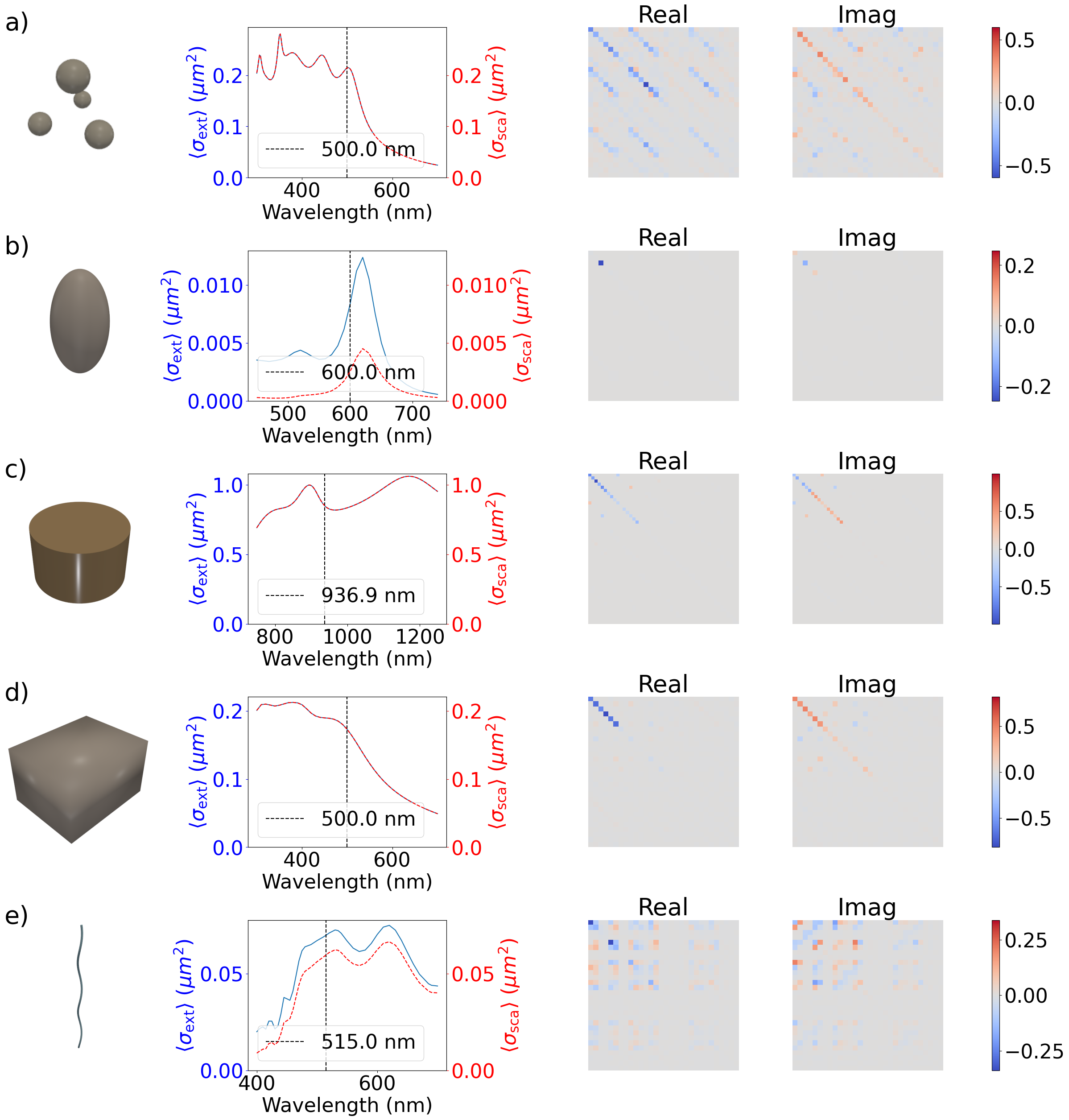}
\caption{Left to right: shape, averaged cross-sections, T-matrix at single wavelength. a) 4 spheres (n = 3) in vacuum: radii: 50, 60, 70, 80 nm, positions:  $a\cdot\left( \left(-\frac{1}{2}, -\frac{\sqrt{3}}{6}, -\frac{\sqrt{6}}{12}\right), \left(+\frac{1}{2}, -\frac{\sqrt{3}}{6}, -\frac{\sqrt{6}}{12}\right),\left(0, \frac{\sqrt{3}}{3}, -\frac{\sqrt{6}}{12}\right),\left(0, 0, \frac{\sqrt{6}}{4}\right) \right)$, $a = 300$ nm
b) Gold~\cite{johnson1972optical} spheroid in water (n = 1.33): $r_{xy}$ = 20 nm, $r_{z}$ = 40 nm
c) TiO$_2$ cylinder (n = 2.5) in vacuum: radius = 250 nm, height = 300 nm
d) Anisotropic cuboid $(\epsilon_{xx}$ = 3.24, $\epsilon_{yy}$ = 4, $\epsilon_{zz}$ = 4.84) in vacuum: L$_x$ = 250 nm, L$_y$ = 200 nm, L$_z$ = 150 nm
e) Silver~\cite{johnson1972optical} helix with right handedness and flat edges in vacuum: number of turns = $2.5$, $r_{\text{wire}} = 21.25$ nm, pitch $= 200$ nm, $r_{\text{helix}} = 41.25$ nm.
}
\label{ExampleCalculations}
\end{figure}

As a demonstration, we selected a few examples of T-matrices computed for different structures. The scripts necessary to generate these files are part of the datasets provided in~\cite{data_format}. These examples are computed using JCMsuite software. The purpose of these examples is to consider them in the future as additional reference examples. If you can reproduce these T-matrices at admissible accuracy with your codes, you likely have an implementation that agrees with the assumptions made throughout the manuscript here. The examples are the following and shown in Fig.~\ref{ExampleCalculations}. The orientation-averaged extinction and scattering cross-sections plotted over a range of wavelengths are of interest here.
a) shows an arrangement of four spheres. This example presents the analytical solution of the scattering problem at optical frequencies. The T-matrix is expanded at the common center of origin, thus a large number of multipoles ($l_\text{max} = 6$) is considered. In b), the spectrum of a gold spheroid in water is demonstrated in the optical frequency range. Maximum mesh size used in the simulations is \SI{1}{\nano \meter} for the spheroid and \SI{3}{\nano \meter} for the embedding medium. This example exhibits rotational symmetry. The material is dispersive and typical plasmonic resonance is captured. Next, in c), the scattering response of a cylinder made of titanium dioxide is computed.  Maximum mesh size used in the simulations is fixed to 6 and \SI{15}{\nano \meter} for object and embedding medium, respectively. The relative permittivity value is taken as a constant. Two peaks at the resonant frequencies are observed. In d), the scatterer geometry does not have rotational symmetry, but mirror symmetry. Artificially set permittivity values were selected to demonstrate the capability of simulating objects made from anisotropic materials. Typical maximum mesh size for the object and the embedding medium were set to $\frac{\lambda_0}{44}$ and  $\frac{\lambda_0}{20}$, where $\lambda_0$ is the vacuum wavelength. As a concluding example, in e) a silver helix is presented, with one dimension considerably larger than the others. Typical maximum mesh size is $\frac{\lambda_0}{80}$,  $\frac{\lambda_0}{20}$ for the object and the embedding medium, correspondingly. No symmetry is present in this geometry.
\newpage
\newpage

\section{Full code of the example in \cref{sec:treams}}
\begin{lstlisting}[language=python,caption=Full script for \textit{treams},label=lst:treams:full]
import h5py
import matplotlib as mpl
import matplotlib.pyplot as plt
import numpy as np
import treams
import treams.io

mpl.rcParams["lines.linewidth"] = 2.5
mpl.rcParams["font.size"] = 20

cyl = treams.io.load_hdf5("cylinder_tio2.tmat.h5")

positions = [[-300, 0, 0], [300, 0, 0]]
cyl_cluster = [
    treams.TMatrix.cluster([tm, tm], positions) for tm in cyl
]
cyl_cluster = [
    tm.interaction.solve().expand(
        treams.SphericalWaveBasis.default(4)
    )
    for tm in cyl_cluster
]

with h5py.File("two_cylinders.tmat.h5", "w") as fobj:
    treams.io.save_hdf5(fobj, cyl_cluster)

lmax = max(cyl[0].basis.l)
cyl_by_lmax = [
    [tm[treams.SphericalWaveBasis.default(i)] for tm in cyl]
    for i in range(1, lmax + 1)
]
cyl_cluster_by_lmax = [
    [tm[treams.SphericalWaveBasis.default(i)] for tm in cyl_cluster]
    for i in range(1, lmax + 1)
]

fig, axs = plt.subplots(1, 2, figsize=(16, 6))

freqs = np.array([tm.k0 * 299792.458 / (2 * np.pi) for tm in cyl])
lambdas = np.array([2 * np.pi / tm.k0 for tm in cyl])
linestyles = [":","--","-.", "-"]
for i in range(lmax):
    axs[0].plot(
        freqs,
        [tm.xs_ext_avg / 1e6 for tm in cyl_by_lmax[i]],
        linestyle=linestyles[i % lmax],
        zorder=-i,
    )

axs[0].set_xlabel("Frequency (THz)")
axs[0].set_ylabel("Cross-section ($\mu m^2$)")
axs[0].set_xlim([240, 400])

ax_top = axs[0].twiny()
ax_top.set_xlabel("Wavelength (nm)")
ax_top.set_xlim(lambdas[0], lambdas[-1])

axs[0].legend([f"$l_\\mathrm{{max}} = {i}$" for i in range(1, lmax + 1)])

for i in range(lmax):
    axs[1].plot(
        freqs,
        [tm.xs_ext_avg / 1e6 for tm in cyl_cluster_by_lmax[i]],
        linestyle=linestyles[i % lmax],
        zorder=-i,
    )

axs[1].set_xlabel("Frequency (THz)")
axs[1].set_ylabel("Cross-section ($\mu m^2$)")
axs[1].set_xlim([240, 400])

ax_top = axs[1].twiny()
ax_top.set_xlabel("Wavelength (nm)")
ax_top.set_xlim(lambdas[0], lambdas[-1])

fig.savefig("xs_cyl_tio2.png")
\end{lstlisting}

\section{
    Alphabetical list of pre-defined \emph{groups}, \emph{datasets} and
    \emph{attributes}
}
\label{sec:list}

The following table contains all reserved names. The color indicates the type:
\emph{groups} are black, \emph{datasets} are \textcolor{blue}{blue}, \emph{attributes}
are \textcolor{red}{red} and \emph{softlinks} are \textcolor{green!50!black}{green}.
Dummy \emph{group} names are in uppercase.
Here, the required entries are printed with an asterisk, the entries with required presence depending on the presence of other entries in the file are denoted with **, while the optional entries, i.e., entries that only provide additional information or have a default value when absent, are listed without an asterisk. If the parent is optional, but has a required child, the child is marked as required.

\begin{footnotesize}
{
\begin{longtable}{  | p{10.6cm} | p{3cm} |  }
\multicolumn{2}{|c|}{Parameter List} \\
Name of the group/dataset& Type \\
\textcolor{red}{inner\_dims} \textrm{for all datasets}  & int \\
/\textcolor{blue}{angular\_frequency} ** & array: float, complex; scalar: float, complex  \\
/\textcolor{blue}{angular\_frequency}/\textcolor{red}{unit} * & str  \\ 
/\textcolor{blue}{angular\_vacuum\_wavenumber}   ** & array: float, complex;  scalar: float, complex   \\
/\textcolor{blue}{angular\_vacuum\_wavenumber}/\textcolor{red}{unit} * & str  \\
/\textcolor{red}{application}  & str  \\
/computation/\textcolor{blue}{analytical\_zeros}  &  array: int \\
/computation/\textcolor{red}{
files} & str  \\
/computation/\textcolor{red}{description} & str  \\
/computation/\textcolor{red}{keywords}  & str  \\
/computation/\textcolor{red}{method} * & str  \\
/computation/method\_parameters/  &   \\
/computation/\textcolor{red}{name}  & str  \\
/computation/\textcolor{red}{software} * & str  \\
/computation/\textcolor{red}{reference}  & str  \\
/\textcolor{red}{description}   & str  \\
/embedding/\textcolor{blue}{bianisotropy}  & array: float, complex  \\
/embedding/\textcolor{blue}{bianisotropy}/\textcolor{red}{coordinate\_system}  & str  \\
/embedding/\textcolor{blue}{chirality}  & array: float, complex;  scalar: float, complex \\
/embedding/\textcolor{blue}{chirality}/\textcolor{red}{coordinate\_system}  & str  \\
/embedding/\textcolor{red}{description}  & str  \\
/embedding/{experimental\_data}/  & \\
/embedding/\textcolor{red}{keywords}  & str  \\
/embedding/\textcolor{red}{name}  & str  \\
/embedding/\textcolor{blue}{non-reciprocity}  & array: float, complex;  scalar: float, complex \\
/embedding/\textcolor{blue}{non-reciprocity}/\textcolor{red}{coordinate\_system}  & str  \\

/embedding/\textcolor{red}{reference}  & str \\
/embedding/\textcolor{blue}{refractive\_index} ** & array: float, complex;  scalar: float, complex  \\
/embedding/\textcolor{blue}{refractive\_index}/\textcolor{red}{coordinate\_system}  & str  \\
/embedding/\textcolor{blue}{relative\_impedance}  ** & array: float, complex;  scalar: float, complex  \\
/embedding/\textcolor{blue}{relative\_permeability} ** & array: float, complex;  scalar: float, complex  \\
/embedding/\textcolor{blue}{refractive\_index}/\textcolor{red}{coordinate\_system}  & str  \\
/embedding/\textcolor{blue}{relative\_permeability}/\textcolor{red}{coordinate\_system}  & str  \\
/embedding/\textcolor{blue}{relative\_permittivity} ** & array: float, complex;  scalar: float, complex  \\
/embedding/\textcolor{blue}{relative\_permittivity}/\textcolor{red}{coordinate\_system}  & str \\

/\textcolor{blue}{frequency} ** & array : float, complex  \\
/\textcolor{blue}{frequency}/\textcolor{red}{unit}  * & str  \\
/\textcolor{red}{keywords}  & str  \\
/\textcolor{green!50!black}{mesh}   & .msh, .STL, etc \\
/modes/\textcolor{blue}{index} ** & array: int \\
/modes/\textcolor{blue}{index\_incident} ** & array: int \\
/modes/\textcolor{blue}{index\_scattered} ** & array: int \\
/modes/\textcolor{blue}{l\_incident} ** & array: int \\
/modes/\textcolor{blue}{l\_scattered} ** & array: int \\
/modes/\textcolor{blue}{l} ** & array: int \\
/modes/\textcolor{blue}{l\_incident} ** & array: int \\
/modes/\textcolor{blue}{l\_scattered} ** & array: int \\
/modes/\textcolor{blue}{m} ** & array: int \\
/modes/\textcolor{blue}{m\_incident} ** & array: int \\
/modes/\textcolor{blue}{m\_scattered} ** & array: int \\
/modes/\textcolor{blue}{polarization} ** & array: str \\
/modes/\textcolor{blue}{polarization\_incident} ** & array: str \\
/modes/\textcolor{blue}{polarization\_scattered} ** & array: str \\
/modes/\textcolor{blue}{positions} ** & array: float \\
/NAME/geometry/\textcolor{blue}{euler\_angles}  & array: float  \\
/NAME/geometry/\textcolor{blue}{expansion\_center}   & array: float  \\
/NAME/geometry/\textcolor{red}{description}   & str  \\
/NAME/geometry/\textcolor{red}{keywords}  & str  \\
/NAME/geometry/mesh/ \textrm{or} /NAME/geometry/\textcolor{blue}{mesh.XYZ} ** (entries containing mesh: same with /computation/ possible ) & .msh, .STL, etc. \\
/NAME/geometry/\textcolor{blue}{mesh.XYZ}/\textcolor{red}{file\_extension}  & str  \\
/NAME/geometry/mesh/\textcolor{red}{unit}  or
/NAME/geometry/\textcolor{blue}{mesh.XYZ}/\textcolor{red}{unit} * & str  \\
/NAME/geometry/\textcolor{red}{name}  & str  \\
/NAME/geometry/\textcolor{blue}{position}   & array: float  \\
/NAME/geometry/\textcolor{red}{shape} ** & str  \\

/NAME/material/\textcolor{blue}{bianisotropy}  & array: float, complex  \\
/NAME/material/\textcolor{blue}{bianisotropy}/\textcolor{red}{coordinate\_system} & str  \\
/NAME/material/\textcolor{blue}{chirality} &  array: float, complex;  scalar: float, complex\\
/NAME/material/\textcolor{blue}{chirality}/\textcolor{red}{coordinate\_system} & str  \\
/NAME/material/\textcolor{red}{description} & str  \\
/NAME/material/\textcolor{blue}{experimental\_data}/  &  \\
/NAME/material/\textcolor{red}{interpolation}  & str  \\
/NAME/material/\textcolor{red}{keywords}  & str  \\
/NAME/material/\textcolor{red}{name}  & str  \\
/NAME/material/\textcolor{blue}{non-reciprocity}  & array: float, complex;  scalar: float, complex \\
/NAME/material/\textcolor{blue}{non-reciprocity}/\textcolor{red}{coordinate\_system}  & str  \\

/NAME/material/\textcolor{red}{reference}   & str \\
/NAME/material/\textcolor{blue}{refractive\_index}  ** & array: float, complex;  scalar: float, complex  \\
/NAME/material/\textcolor{blue}{refractive\_index}/\textcolor{red}{coordinate\_system}  & str  \\
/NAME/material/\textcolor{blue}{relative\_impedance}  ** & array: float, complex;  scalar: float, complex  \\
/NAME/material/\textcolor{blue}{relative\_impedance}/\textcolor{red}{coordinate\_system} & str  \\
/NAME/material/\textcolor{blue}{relative\_permeability} ** & array: float, complex;  scalar: float, complex  \\
/NAME/material/\textcolor{blue}{relative\_permeability}/\textcolor{red}{coordinate\_system}  & str  \\
/NAME/material/\textcolor{blue}{relative\_permittivity} ** & array: float, complex;  scalar: float, complex  \\
/NAME/material/\textcolor{blue}{relative\_permittivity}/\textcolor{red}{coordinate\_system}  & str \\
/\textcolor{blue}{rmatrix}  & array: float, complex \\
/\textcolor{red}{storage\_format\_version} *  & str  \\ 
/\textcolor{blue}{tmatrix} * & array: float, complex\\
/\textcolor{blue}{vacuum\_wavelength} ** & array: float, complex;  scalar: float, complex  \\
/\textcolor{blue}{vacuum\_wavelength}/\textcolor{red}{unit} *  & array: float, complex;  scalar: float, complex  \\
/\textcolor{blue}{vacuum\_wavenumber} ** & array: float, complex;  scalar: float, complex  \\
/\textcolor{blue}{vacuum\_wavenumber}/\textcolor{red}{unit}  * & str  \\ 
\hline
\end{longtable}
}
\end{footnotesize}
\bibliographystyle{elsarticle-num} 
\bibliography{biblio}
\end{document}